\pdfoutput=1
\documentclass[11pt]{article}

\usepackage[preprint]{acl}

\usepackage[T1]{fontenc}
\usepackage[utf8]{inputenc}

\usepackage{microtype}

\usepackage{inconsolata}

\usepackage{graphicx}
\usepackage{times}
\usepackage{latexsym}
\usepackage{latexsym}
\usepackage{booktabs}  % 为\toprule等命令
\usepackage{amssymb}  % 为\checkmark命令
\usepackage{pifont}
\usepackage{subcaption}
\usepackage{multirow}
\usepackage{colortbl}
\usepackage{xcolor}

\title{\textsc{Muse}: A Multimodal Conversational Recommendation Dataset with Scenario-Grounded User Profiles}

\author{Zihan Wang$^1$, Xiaocui Yang$^1$, Yongkang Liu$^1$, Shi Feng$^{1,\dag}$ \\ 
        \textbf{Daling Wang}$^1$, \textbf{Yifei Zhang}$^1$\\
       $^1$School of Computer Science and Engineering, Northeastern University, Shenyang, China \\
       {\normalsize \texttt{2310744@stu.neu.edu.cn}}, ~~
       {\normalsize \texttt{yangxiaocui@cse.neu.edu.cn}}, ~~
       {\normalsize \texttt{misonsky@163.com}} \\
       {\normalsize \texttt{\{fengshi, wangdaling, zhangyifei\}@cse.neu.edu.cn}}
       }
\begin{document}
\maketitle

\begin{abstract}
Current conversational recommendation systems focus predominantly on text. 
However, real-world recommendation settings are generally multimodal, causing a significant gap between existing research and practical applications. 
To address this issue, we propose \textsc{Muse}, the first multimodal conversational recommendation dataset. 
\textsc{Muse} comprises 83,148 utterances from 7,000 conversations centered around the Clothing domain. 
Each conversation contains comprehensive multimodal interactions, rich elements, and natural dialogues. 
Data in \textsc{Muse} are automatically synthesized by a multi-agent framework powered by multimodal large language models (MLLMs). 
It innovatively derives user profiles from real-world scenarios rather than depending on manual design and history data for better scalability, and then it fulfills conversation simulation and optimization. 
Both human and LLM evaluations demonstrate the high quality of conversations in \textsc{Muse}. 
Additionally, fine-tuning experiments on three MLLMs demonstrate \textsc{Muse}'s learnable patterns for recommendations and responses, confirming its value for multimodal conversational recommendation. 
Our dataset and codes are available at \url{https://anonymous.4open.science/r/Muse-0086}.  
\end{abstract}

\def\thefootnote{\dag}\footnotetext{Corresponding author.}

\section{Introduction}
Conversational recommendation (CR) \cite{lei2020conversational} is an emerging research field. 
It leverages natural language to deliver personalized, context-aware suggestions for users.
Unlike the traditional implicit recommendation paradigm \cite{jalili2018evaluating,wang2019sequential,wang2021survey}, CR emphasizes both the recommendation performance and the real-time dialogue with users. 
Some existing CR datasets supporting the research, such as Redial \cite{li2018towards} and TG-Redial \cite{zhou2020topic}, are launched through crowdsourcing. 
Crowdworkers take on dual roles as users and rec-assistants, interacting with each other to generate CR data. 
Innovatively, Pearl \cite{kim2024pearl} and LLM-Redial \cite{liang2024llmredial} harness the advanced capabilities of large language models (LLMs) \cite{chang2024survey} to fulfill the simulation of conversation. 

\begin{figure}[t]
 \centering
 \includegraphics[width=0.45\textwidth]{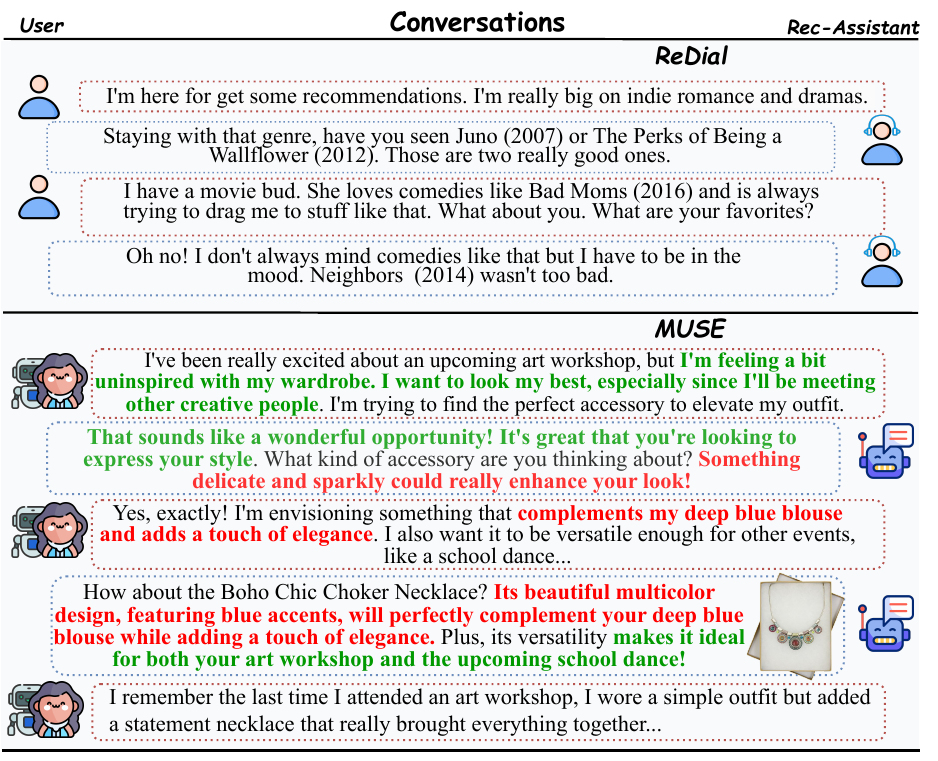}
 \vspace{-2mm}
 \caption{Comparison of data cases from Redial and \textsc{Muse}. 
 Red denotes interactions about visual features, and green shows scenario-related content.}
 \label{fig:example}
 \vspace{-5mm}
\end{figure}

\begin{table*}[htbp]
\centering
\small
\caption{Comparison Between \textsc{Muse} and some dialogue datasets. \textbf{Exp. Rec} means `Expainable Recommendation.' \textbf{Conv. Scal.} and \textbf{U/I. Scal.} are the scalability of conversations and users/items}
\vspace{-1mm}
\label{tab:overall_compare}
\resizebox{0.93\textwidth}{!}{
\begin{tabular}{@{}llllllllll@{}}
\toprule
\textbf{Datasets} & \textbf{\#Dial.} & \textbf{\#Utter.} & \textbf{Domains} & \textbf{Types} & \textbf{Exp. Rec
} & \textbf{Conv. Scal.} & \textbf{U/I. Scal.} & \textbf{U. Profile} & \textbf{Modal} \\
\midrule
\multicolumn{10}{l}{\hspace{13em}\textit{\textbf{Multimodal Dialogue Datasets}}} \\
\midrule
\rowcolor{yellow!4} SURE \cite{long2023multimodal} & 12K & 223K & Fashion, furniture & Task-oriented.Dial & $-$ & $\times$ & $-$ & $-$ & VR+Text \\
\rowcolor{yellow!4} SIMMC1.0 \cite{crook2021situated} & 13K & 169K & Fashion, furniture & Task-oriented.Dial & $-$ & $\times$ & $-$ & $-$ & VR+Text \\
\rowcolor{yellow!4} SIMMC2.0 \cite{kottur2021simmc} & 11K & 117K & Fashion, furniture & Task-oriented.Dial & $-$ & $\times$ & $-$ & $-$ & VR+Text \\
\rowcolor{yellow!4} IGC \cite{mostafazadeh2017image} & 4K & 25K & Image concepts & Image-based QA & $-$ & $\times$ & $-$ & $-$ & Image+Text \\
\rowcolor{yellow!4} GuessWhat \cite{de2017guesswhat} & 155K & 1.6M & Image concepts & Image-based QA & $-$ & $\times$ & $-$ & $-$ & Image+Text \\
\rowcolor{yellow!4} MMD \cite{saha2018mmd} & 150K & 6M & Fashion & Conv.Search & $-$ & $\times$ & $-$ & $-$ & Image+Text \\
\rowcolor{yellow!4} MMCONV \cite{liaoC21mmconv} & 5.1K & 39.7K & Travel & Conv.Search & $-$ & $\times$ & $-$ & $-$ & Image+Text \\
\midrule
\multicolumn{10}{l}{\hspace{13em}\textit{\textbf{Conversational Recommendation Datasets}}} \\
\midrule
\rowcolor{blue!4} Redial \cite{li2018towards} & 10K & 182K & Movie & Conv.Rec & $\times$ & $\times$ & $\times$ & From human design & Text \\
\rowcolor{blue!4} OpenDialKG \cite{moon2019opendial} & 15K & 91K & Movie, Book & Conv.Rec & $\times$ & $\times$ & $\times$ & From human design & Text \\
\rowcolor{blue!4} TG-Redial \cite{zhou2020topic} & 10K & 129K & Movie & Conv.Rec & $\times$ & $\times$ & $\times$ & From human design & Text \\
\rowcolor{blue!4} DuRecDial \cite{liu2020durec} & 10.2K & 156K & Movie, Music, Food & Conv.Rec & $\times$ & $\times$ & $\times$ & From human design & Text \\
\rowcolor{blue!4} INSPIRED \cite{hayati2020inspired} & 1K & 35K & Movie & Conv.Rec & $\times$ & $\times$ & $\times$ & From human design & Text \\
\rowcolor{blue!4} Pearl \cite{kim2024pearl} & 57.2K & 482K & Movie & Conv.Rec & $\checkmark$ & $\checkmark$ & $\times$ & From history data & Text \\
\rowcolor{blue!4} LLM-Redial \cite{liang2024llmredial} & 47.6K & 548K & Movie, Book, Sports & Conv.Rec & $\checkmark$ & $\checkmark$ & $\times$ & From history data & Text \\
\midrule
\textsc{Muse} & 7K & 83K & Cloth, Shoes, Jewelry & Conv.Rec & $\checkmark$ & $\checkmark$ & $\checkmark$ & From real-world scenarios& Image+Text \\
\bottomrule
\end{tabular}
}
\vspace{-5mm}
\end{table*}
While these datasets have significant contributions, they still possess limitations. 
(1) These datasets are predominantly limited to textual modality. 
However, like other recommendation fields \cite{zhou2023comprehensive}, text-only data is insufficient to simulate the multisensory decision-making processes that characterize real-world shopping behaviors. 
Multimodal information is particularly crucial for visually driven fields, such as clothing and food. 
(2) Their scalability is limited. 
Given CR's data-driven nature, CR datasets are supposed to be scaled to include more conversations and a wider range of users and items \cite{liang2024llmredial}, enabling the development of more comprehensive CR systems. 
Existing LLM-based methods have demonstrated the ability to scale conversation volume, which partially addresses the limitations of crowdsourcing datasets. 
However, due to their heavy reliance on user history data, their user and item coverage remains confined to historically collected data. 
Moreover, facing increasingly stringent privacy regulations \cite{voigt2017eu,regulation2016regulation,harding2019understanding} and cold-start situations \cite{lam2008addressing}, these methods become hard to apply. 

To tackle these challenges, we introduce \textsc{Muse}, a \textbf{MU}ltimodal Conversational Recommendation Dataset with \textbf{S}c\textbf{E}nario-grounded user profiles. 
To the best of our knowledge, it is the first multimodal dialogue dataset specifically designed for CR tasks. 
\textsc{Muse} is based on real-world products from the multimodal dataset, Amazon Cloth, Shoes, and Jewelry \cite{hou2024bridging}, and comprises a total of 7,000 multimodal conversations. 
Figure.~\ref{fig:example} illustrates comparative case studies of conversations drawn from ReDial and \textsc{Muse}.
Motivated by the effectiveness of LLM-based data synthesis in previous works, a multi-agent framework powered by Multimodal LLMs (MLLMs) facilitates conversations in \textsc{Muse}. 
The framework has three modules: Scenario-Grounded User Profile Generator, Simulated Conversation Generator, and Conversation Optimizer. 
Inspired by the understanding that user engagement in recommendations involves both preference-based interactions (“I like...”) and scenario-grounded requirements \cite{paul2016predicting} (“Need for wedding/sports...”), the Scenario-Grounded User Profile Generator adopts an innovative approach. 
It places roles within different real-world scenarios and identifies their diverse needs to match different suitable target items, resulting in scenario-grounded user profiles. 
The infinite diversity of real-world scenarios naturally brings the scalability of both users and items. 
The Simulated Conversation Generator utilizes an ensemble of interconnected sub-agents that synergistically leverage multimodal information for advanced dialogue simulation. 
The system integrates fine-grained multimodal characteristics into the conversation process and adds natural chit-chat parts that simulate human conversations. 
The Conversation Optimizer consists of a Rewriter and a Reviewer. 
The former amplifies dialogue diversity through both sentence/word variation and colloquial elements, and the latter filters out conversations that are not eligible, ensuring data quality in \textsc{Muse}. 
As a result, \textsc{Muse} offers diverse element-rich multimodal conversations and keeps remarkable scalability because of the framework. 
Table.~\ref{tab:overall_compare} presents a comparison between \textsc{Muse} and representative datasets, which are categorized into multimodal dialogue datasets and CR datasets. 

We perform comprehensive assessments of \textsc{Muse} conversations through both human evaluation and LLM analysis, evaluating them from both global and granular perspectives. 
Empirical results indicate that \textsc{Muse} generates dialogues with exceptional fluency, diversity, depth of bilateral interaction, and multimodal coherence. 
To validate the utility of \textsc{Muse} as a multimodal conversational recommendation (MCR) dataset, we conduct extensive evaluations on three representative open-source MLLMs under both zero-shot and fine-tuned configurations. 
The quantitative results demonstrate \textsc{Muse}'s capacity to facilitate reliable recommendation reasoning and response generation for CR, establishing its value as a benchmark dataset for MCR. 

\section{Related Work}
\subsection{Conversational Recommendation}
Conversational recommendation (CR) research can be broadly divided into two categories \cite{jannach2021survey,fu2020tutorial,jannach2022conversational}.
The first frames the task as a multi-step decision-making process, leveraging reinforcement learning to minimize the number of conversation rounds required to identify the target item \cite{deng2021unified,zhang2022multiple}.
The second prioritizes natural language communication, aiming to gain a deep understanding of user preferences through conversation and, in some cases, even influence those preferences \cite{li2018towards,wang2022towards,ravaut2024parameter}.
Our work centers on the latter. 
Since the inception of the Redial dataset \cite{li2018towards}, numerous crowdsourcing datasets \cite{moon2019opendial,zhou2020topic,liu2021durecdial,hayati2020inspired} have emerged, expanding the CR task across various data. 
Pearl \cite{kim2024pearl} and LLMRedial \cite{liang2024llmredial} introduce innovative approaches with LLMs. 
However, these datasets are limited to plain text. 
\textsc{Muse} marks a major breakthrough in the field as a multimodal conversational recommendation dataset. 

\subsection{LLM-Driven Data Synthesis}
Large Language Models (LLMs) possess extensive and diverse world knowledge \cite{zhao2023survey}, enabling them to comprehend and generate complex language with human-like proficiency. 
Therefore, LLMs have demonstrated remarkable potential in data synthesis \cite{ding-etal-2024-data,wang2022self,sahu2022data,liu2024unified}, especially in generating conversational datasets \cite{abbasiantaeb2024let,kim2022soda}. 
In the realm of conversational recommendation, two prominent approaches—Pearl \cite{kim2024pearl} and LLM-Redial \cite{liang2024llmredial}—have gained recognition. 
While the two approaches enable increased dialogue quantity, they struggle to diversify user and item coverage due to their reliance on history data. 
\textsc{Muse} ingeniously harnesses the vast world knowledge embedded in multimodal LLMs to craft user profiles and match target items, enabling full scalability in conversational recommendation datasets.

\section{\textsc{Muse} Construction} 
\begin{figure*}[t]
 \centering
 \includegraphics[width=0.95\textwidth]{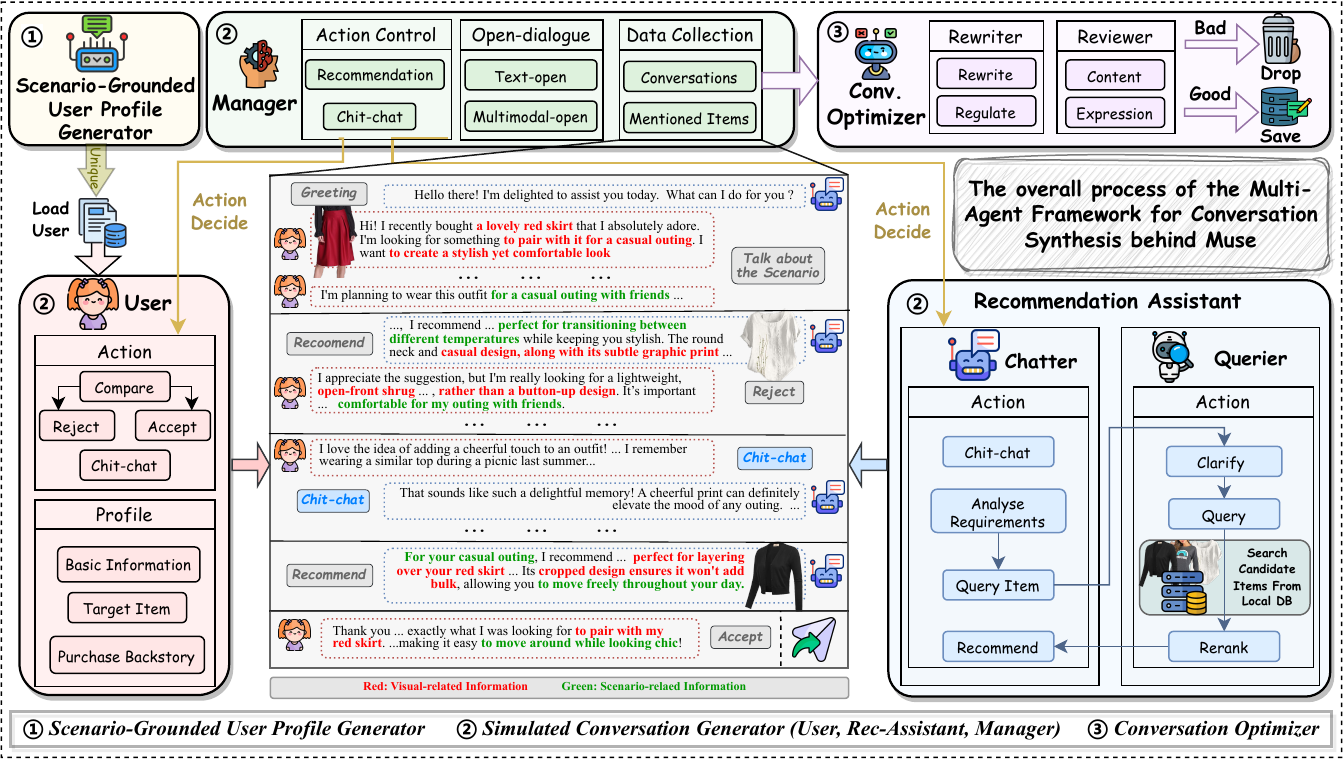}
 \caption{The multi-agent framework for synthesizing MCR data in \textsc{Muse}.}
 \vspace{-2mm}
 \label{fig:Muse}
 \vspace{-4mm}
\end{figure*}

We construct \textsc{Muse}, the first MCR dataset, which is built based on real-world clothing product information. 
In this section, we introduce the multi-agent framework behind \textsc{Muse} for conversation synthesis, which is organized into three functional components as descriptions in Figure.~\ref{fig:Muse}: \ding{172} Scenario-Grounded User Profile Generator; \ding{173} Simulated Conversation Generator; \ding{174} Conversation Optimizer. 
The main backbone LLM of this framework is \texttt{gpt-4o-mini}\footnote{https://openai.com}; check Appendix \ref{appendix:mmsetting} for more MLLM settings.

\subsection{Data Preprocess}
To anchor \textsc{Muse} conversations in real-world products, we utilize the Amazon Clothing, Shoes, and Jewelry dataset \cite{hou2024bridging}, which combines both textual and visual information.
Using multimodal product information, we build a local product database to support subsequent product retrieval operations, where each product is accompanied by a main image and a text description.  
The detailed processing is documented in the Appendix \ref{appendix:data_preprocess}. 

\subsection{Scenario-Grounded User Profile Generator}
\begin{figure}[t]
 \centering
 \includegraphics[width=0.46\textwidth]{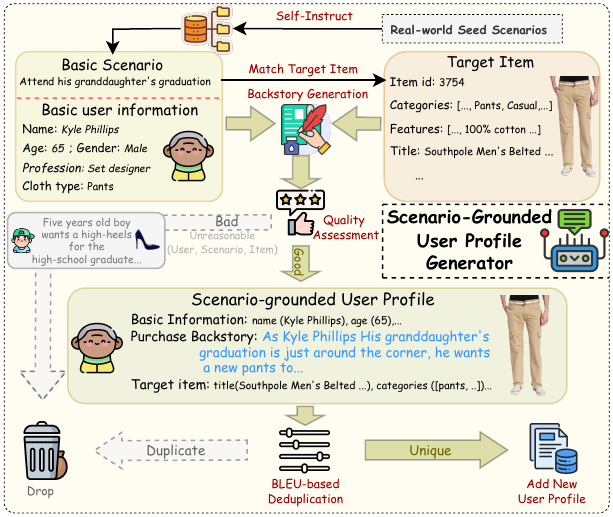}
 \vspace{-2mm}
 \caption{Workflow of the scenario-grounded user profile generator }
 \label{fig:scenario-generator}
 \vspace{-5mm}
\end{figure}

In the real world, users' immediate purchasing decisions are not solely driven by personal interests but are significantly influenced by a multitude of external factors \cite{piligrimiene2020internal}.
Among these, scenario context plays a pivotal role, as different scenarios, such as buying a suit for a party or a T-shirt for the summer, drive diverse consumer needs, requiring tailored products to meet them. 
Note that these requirements can correspond to multimodal product features. 
Ideally, every user need can be traced back to a real-life scenario, just as suitable scenarios and users can be identified for any product, which naturally enables flexible scalability in both user and item. 
Building on this concept, we develop the Scenario-Grounded User Profile Generator, whose entire process can be divided into two steps, each assigned to a dedicated agent. 
The first step is to collect diverse real-world basic scenarios. 
In the second step, we situate users in various scenarios and match them with products that align with both the scenario requirements and their individual characteristics to get detailed user profiles to support the subsequent user simulator. 
Figure.~\ref{fig:scenario-generator} illustrates the whole workflow. 

\subsubsection{Basic Scenario Generation}
Basic scenarios reflect real-world events that shape user shopping behavior.
To capture a diverse range of such scenarios, the Basic Scenario Generator harnesses the expansive capabilities of LLMs.
It begins with a set of seed scenarios related to clothing purchases, including but not limited to attending important occasions, meeting athletic needs, celebrating special dates, and purchasing gifts for friends and family. 
Utilizing the self-instruct method \cite{wang2022self}, we expand the scenarios. 
During this expansion, the BLEU \cite{papineni-etal-2002-bleu} metric is adopted to eliminate duplicates, ensuring that the collected real-world scenarios maintain their diversity. 
Ultimately, we identify 593 basic scenarios for \textsc{Muse}. 

\subsubsection{User Profile Generation} 
In our design, a complete user profile consists of three key components: basic user information, target products, and the purchase backstory. 
To enhance the traditional user profile, we incorporate detailed driving information about the user's current shopping behavior, which we call the Purchase Backstory.
This addition provides a more comprehensive explanation of the user's motivation based on scenario requirements, enabling more accurate role simulation \cite{chen2024oscars}. 
Specifically, we first generate basic user information, including demographic details such as age and occupation, and then match the user with a specific scenario and a target product. 
In this step, we utilize MLLMs to evaluate the rationality of the (user, scenario, product) combination from two perspectives: user-product matching and scenario-product matching and screen low-quality results. 
In the second step, the MLLMs generate a purchase backstory for each reasonable combination (user, scenario, product). 
Here, we also employ the BLEU metric to remove duplicates, ensuring the uniqueness of each purchase backstory. 
By integrating the above information, we create scenario-grounded user profiles. 

\subsection{Simulated Conversation Generator}
This generator comprises three specialized agents: a User Simulator and a Rec-assistant Simulator, both dedicated to simulating user dialogues through iterative multimodal interactions, and a Manager that oversees the actions of the two simulators and collects data, as shown in Figure.~\ref{fig:Muse}. 

\subsubsection{User Simulator}
The user simulator adopts scenario-grounded user profiles generated above for role-playing. 
Research by \cite{zhang2024generative,wang2023large} demonstrates that advanced LLMs are highly effective in performing user simulation tasks. 
Among the factors in the provided user profiles, two stand out as having the greatest impact on CR. 
(1) \textbf{Scenario-based Requirements} represent the users' primary concerns and needs, which are seamlessly incorporated into open-dialogues. 
Open-dialogues represent the first few rounds of the conversation, in which rec-assistant learns some basic requirements from the user through question and answer, explained in Section \ref{agent:manager}. 
(2) \textbf{Target requirements} specify the desired features of the target item of the user. 
These requirements align with Pearl's methodology \cite{kim2024pearl}, equipping the user simulator with a clear framework to generate more precise and insightful feedback. 

The user simulator is designed to perform two primary types of actions. 
(1) \textbf{Actions towards recommended items.} 
The user compares the recommended product's visual and textual information against his/her requirements to identify any discrepancies. 
If the product falls short of the user's needs, the simulator provides a logical justification for rejection. 
When the product aligns with requirements, it generates an acceptance response with appropriate appreciation. 
(2) \textbf{Chit-chat.} \cite{wang2023rethinking,liang2024llmredial} emphasize chit-chat is a vital component of natural human dialogue. 
Existing LLM-based datasets have largely overlooked it while \textsc{Muse} acknowledges and incorporates this crucial aspect into conversations. 

\subsubsection{Rec-assistant Simulator}
The rec-assistant simulator is composed of two sub-agents: \textbf{Chatter}, which specializes in user communication, and \textbf{Querier}, which handles recommendations and provides product information. 
This division is designed to align with the primary goals of conversational recommendations: delivering high-quality interactions and offering products that meet user expectations. 

\textbf{Chatter.} 
Chatter prioritizes the quality of its responses and supports two main actions: recommendation and chit-chat.
For recommendations, it prompts the Querier to provide a suitable product. Based on the contextual information, Chatter evaluates the compatibility between the product's multimodal information and the user's needs, using this alignment as a key selling point in its recommendation.
Chit-chat focuses on responding to the user's casual dialogues, offering engaging content and emotion support for better user experiences.

\textbf{Querier.} 
Querier is responsible for finding products that meet the user's exposed requirements. 
First, the Querier analyzes the overall current conversation to craft the user's interests. 
Then, it clarifies the user's interests because in natural language expression, the user's preferences may be vague, and they need to be matched with the local database that focuses on attribute descriptions.  
For example, "need a quick-drying clothing" will be clarified as "need clothes made of polyester, modal... materials." 
Based on the clarified user needs, a rough retrieval is performed from local products, followed by LLM-powered reranking to identify the best-matched product. 
Chatter is provided with multimodal information of the best-matched product. 
If the round limit is reached, the Querier provides the user's target product to end the conversation. 

\subsubsection{Manager}
\label{agent:manager}
The Manager's responsibilities encompass three key functions: (1) initiating the open-dialogues to start conversations, (2) orchestrating action control to regulate the exchanges between users and the rec-assistant in each round, and (3) performing data collection to document the conversation content. 

\textbf{Open-dialogue.}
Open-dialogues refer to some initial exchanges between the user and the system at the start of the recommendation process, primarily involving greetings and basic inquiries.  
It is commonly observed in artificial CR datasets \cite{li2018towards,zhou2020topic}, underscoring its significance as a feature of human conversations. 
Therefore, in \textsc{Muse}, we integrate the simulation of two types open-dialogue to further imitate human conversations. 
One is the text-open, like those text-only datasets. 
The other is the multimodal-open, designed to accommodate users who need to express their needs with images. 
We constrain multimodal-open dialogues specifically to outfit-matching applications, which is a common concern in Clothing data \cite{lin2019explainable,xu2024difashion}. 
In the context of clothing datasets, outfit coordination represents a typical multimodal-open use case - such as "Please select a T-shirt to complement these pants" accompanied by an image. 
More details are in Appendix \ref{appendix:open-dialogue}. 

\textbf{Action Control.}
The Manager needs to guide the actions of the user and rec-assistant in rounds, mainly to control the distribution of chit-chat rounds and recommendation rounds. 
Different action monitoring strategies can be tailored to specific contexts. 
For example, "the longer the conversation, the less likely users are to engage in chit-chat," which is the approach we have adopted. 

\textbf{Data Collection.}
The process is straightforward: Manager records and organizes the conversation content in a structured format. 

\subsection{Conversation Optimizer}
The conversation optimization system consists of two specialized agents—the Rewriter and Reviewer—who work in tandem to enhance conversation diversity and perform conversation quality assessments to screen low-quality conversations. 

\subsubsection{Rewriter}
In the previous conversation generation process, to ensure the agents strictly followed the instructions in the prompt, maintain dialogue stability, and relieve LLM hallucinations, we set the MLLM model temperature for both the User and Rec-Assistant Simulators to 0.1. 
However, the setting results in the sentence structure and wording of the User and Rec-Assistant outputs being relatively repetitive. 
To introduce more diverse expressions in \textsc{Muse}, we implement a Rewriter tasked with modifying the sentence structure and wording of the dialogue. 
An internal supervision mechanism is employed to ensure consistency in content, with particular emphasis on preserving the accuracy and immutability of product attribute information. Additionally, we incorporate probabilistic "use colloquial expressions" instructions into the prompt to generate more human-like responses. 
As a result, greater diversity in expressions is achieved while maintaining logical coherence. 
The details are in Appendix \ref{appendix:rewriter_structure}. 

\subsubsection{Reviewer}
After completing all conversation rounds, we evaluate the overall quality of the conversation to filter out low-quality products. 
To ensure a reliable assessment, we use three key indicators: content quality, logical fluency, and user consistency. 
These indicators collectively assess the conversation content, and a scoring strategy is applied to calculate the total score. 
The combined results are then used to evaluate and screen the conversations effectively. 

% \subsection{Dialogue Filtering}
% In the previous process, in order to stabilize the quality and diverisity our framework actually conducted three automatic screenings. 
% (1) BLEU deduplication in basic scenario. 
% (2) Quality screening based on the matching rationality of users, scenarios, and products. 
% (3) BLEU deduplication in generated user backstory. 
% (4) Quality screening based on Reviewer's scoring of the whole conversation.
% Considering the evaluation of the overall reliability of the public dataset, we also use manual screening, but it turn out that after the previous four screenings, the quality of the retained conversations was already quite guaranteed. 
% Manual screening only filter out a small number of unqualified conversations discussed in the experimental section. 

\section{Experiments}
In this section, we present comprehensive experiments to validate the value of \textsc{Muse}. 
Initially, we analyze various data parameters in \textsc{Muse} along with the composition of its dialogue elements. 
Then, we assess the dialogue quality in \textsc{Muse} using a two-fold evaluation methodology: overall conversation-level quality and utterance-level quality. 
Furthermore, to demonstrate the practical utility of our dataset for the conversational recommendation, we conduct experiments with three representative open-source MLLMs, evaluating their performance in both recommendation generation and response generation on \textsc{Muse}. 

\begin{table}[htbp]
\centering
\caption{Dataset Statistics of \textsc{Muse} and other datasets. }
\vspace{-1mm}
\resizebox{0.45\textwidth}{!}{
\begin{tabular}{llllll}
\hline
\textbf{Metric} & \textbf{MMCONV} & \textbf{Redial} & \textbf{INSPIRED} & \textbf{Pearl} & \textsc{\textbf{Muse}} \\ 
\hline
\#Users & $-$ & 1.0K & 1.0K & 4.7K & 7.0K \\
\#Items & $-$ & 51.7K & 1.7K & 9.4K & 13.7K\\
\#Images & 114K & $-$ & $-$ & $-$ & 13.7K \\
\midrule
\#4-Grams & 230K & 38K & 140K & 3.5M & 2.3M \\
Distinct-3 & 0.24 & 0.27 & 0.55 & 0.09 & 0.30 \\
Distinct-4 & 0.38 & 0.48 & 0.76 & 0.18 & 0.54 \\
Avg.word/Turn & 12.8 & 7.6 & 7.9 & 34.7 & 46.6 \\ 
\hline
\end{tabular}
}
\vspace{-5mm}
\label{tab:dataset_stats}
\end{table}

\begin{figure}
    \centering
    \includegraphics[width=0.9\linewidth]{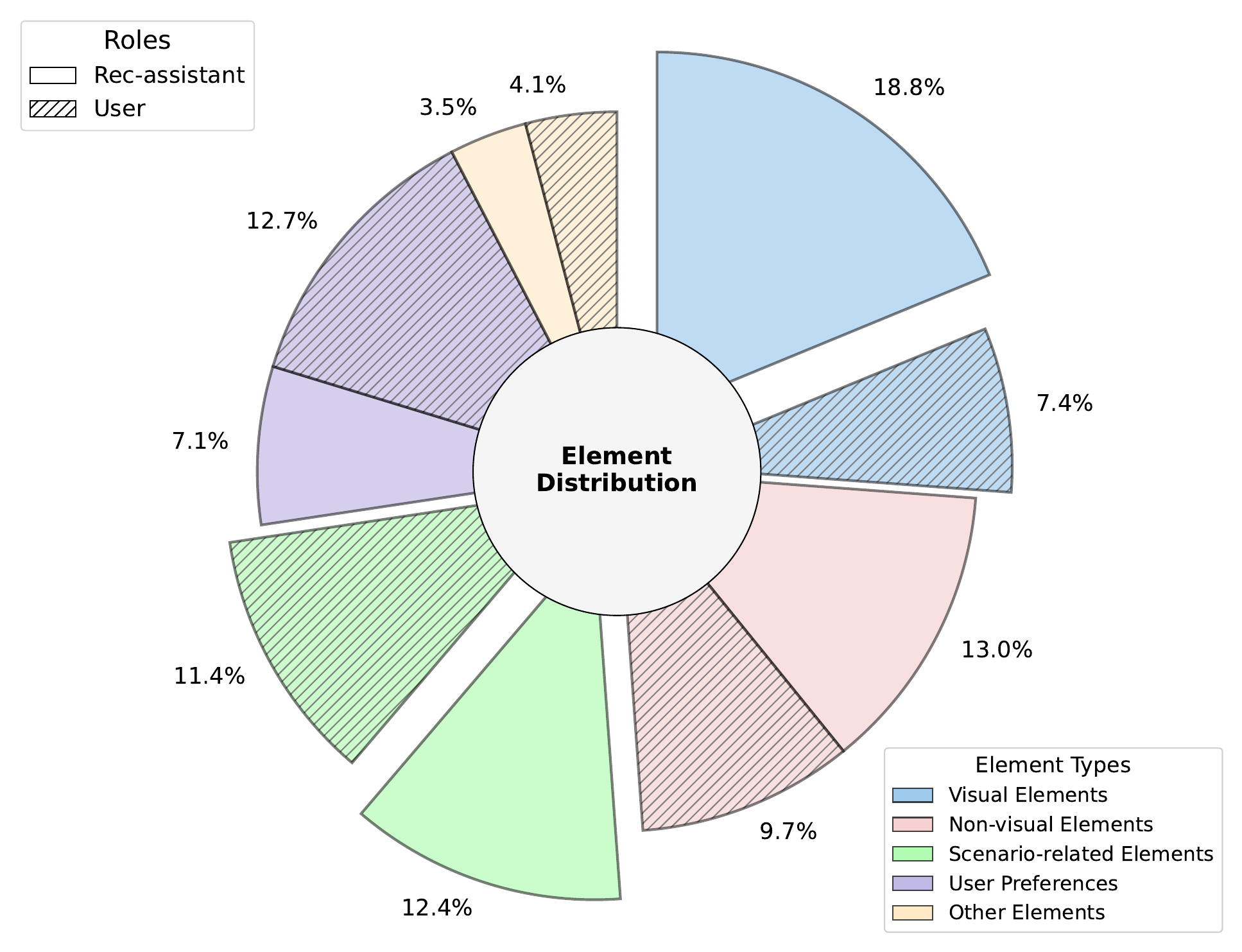}
    \vspace{-2mm}
    \caption{Distribution of dialogue Elements in \textsc{Muse}}
    \label{fig:composition}
    \vspace{-5mm}
\end{figure}

% \begin{table}
% \centering
% \caption{Comparison of different datasets across multiple evaluation metrics}
% \vspace{-1mm}
% \begin{tabular}{lccccc}
% \toprule
% \textbf{Metrics} & \textbf{MMCONV} & \textbf{Redial} & \textbf{\textsc{Pearl}} & \textbf{\textsc{Muse}} \\
% \midrule
% \textbf{Natural}(0-2) & 1.41 & 1.57 & 1.66 & \textbf{1.85} \\
% \textbf{Logical}(0-2) & 1.41 & 1.60 & 1.78 & \textbf{1.88} \\
% \textbf{Informative}(0-2) & 1.37 & 1.53 & 1.67 & \textbf{1.80} \\
% \textbf{P-C Correlation}(0-2) & 1.68 & 1.92 & 1.50 & \textbf{1.98} \\
% \textbf{I-T Correspondence}(0-2) & 1.35 & - & - & \textbf{1.91} \\
% \bottomrule
% \end{tabular}
% \vspace{-3mm}
% \label{tab:conversation-level quality}
% \end{table}

\subsection{Statistics of \textsc{Muse}}
A comparative analysis of basic statistics between \textsc{Muse} and three conversational recommendation datasets is shown in Table.~\ref{tab:dataset_stats}. 
Both \textsc{Pearl} and \textsc{Muse} demonstrate notably higher average word counts per conversation. 
It indicates that synthesized data from LLMs tend to produce more extensive expressions. 
Furthermore, the higher 4-grams \cite{loper2002nltk} (nltk==3.9.1) confirms the presence of more distinctive content, more likely due to the detailed articulation of fine-grained product features. 
An intriguing observation emerges: \textsc{Pearl} exhibits notably low distinct-n \cite{li2015diversity}. 
Upon further investigation, we discover that while Pearl contains richer product content, it demonstrates excessive repetition in its dialogue patterns and phrasal expressions. 
The presence of rigid sentence patterns can substantially reduce conversational diversity and hinder the generalization ability of trained systems. 
\textsc{Muse} overcomes this limitation by employing a Rewriter, which effectively diversifies sentence structures and lexical choices, thereby improving overall data quality. 
More discussion is placed in Appendix \ref{appendix:n_gram}. 

Also, we employ \texttt{gpt-4o} to extract and classify dialogue elements across 5,000 dialogues in \textsc{Muse} to conduct an in-depth analysis. 
Figure.~\ref{fig:composition} illustrates the distribution of main dialogue elements in the Muse dataset. 
The dialogue elements are systematically categorized based on both their functional types (Element Types) and participant sources (Roles). 
The analysis reveals that both users and Rec-assistants frequently incorporate visual and scenario-related elements in their exchanges, accounting for a substantial proportion of the dialogue content.
This highlights the importance of these elements in enriching interactions and supporting communication within \textsc{Muse}. 

\begin{table}
\centering
\caption{Comparison of different datasets across multiple evaluation metrics}
\vspace{-2mm}
\resizebox{0.45\textwidth}{!}{
\begin{tabular}{lcccccc}
\toprule
\textbf{Metrics} & \textbf{MMCONV} & \textbf{Redial} & \textbf{INSPIRED} & \textbf{\textsc{Pearl}} & \textbf{\textsc{Muse}} \\
\midrule
\textbf{Natural}(0-2) & 1.41 & 1.57 &1.71 & 1.66 & \textbf{1.85} \\
\textbf{Logical}(0-2) & 1.41 & 1.60 & 1.62  &1.78 & \textbf{1.88} \\
\textbf{Informative}(0-2) & 1.37 & 1.53 &1.51 & 1.67 & \textbf{1.80} \\
\textbf{P-C Correlation}(0-2) & 1.68 & 1.92 &1.83& 1.50 & \textbf{1.98} \\
\textbf{I-T Correspondence}(0-2) & 1.35 & - & - & - & \textbf{1.91} \\
\bottomrule
\end{tabular}
}
\vspace{-2mm}
\label{tab:conversation-level quality}
\end{table}
\begin{figure}
    \centering
    \includegraphics[width=0.9\linewidth]{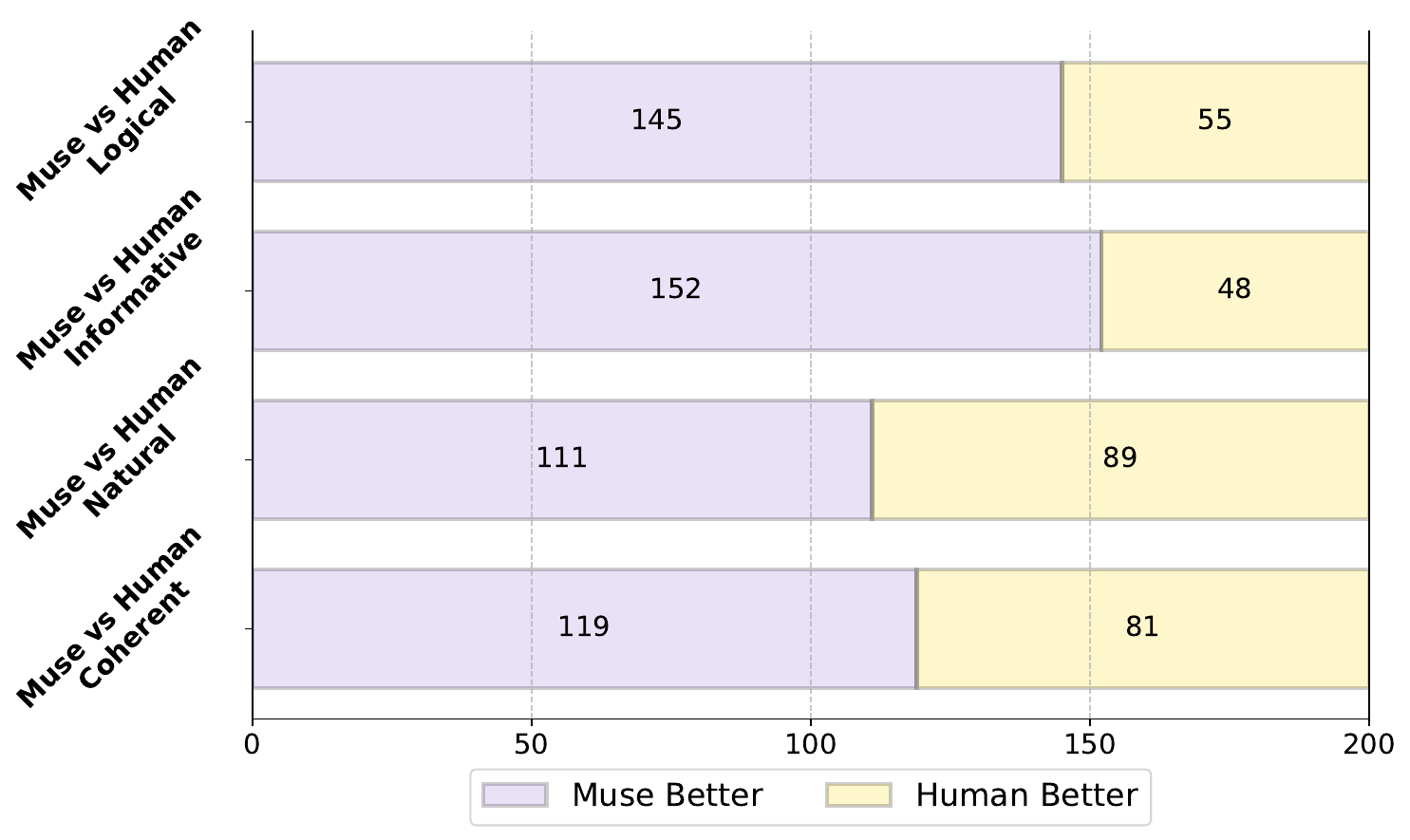}
    \vspace{-2mm}
    \caption{Utterance-level comparison: the quality between human responses and Muse's utterances, }
    \label{fig:utterance-level}
    \vspace{-5mm}
\end{figure}

\subsection{Conversation-level Evaluation} 
To evaluate the global quality of our dataset, we conduct comparative analyses against four representative datasets with similar characteristics to \textsc{Muse}: MMCONV, Redial, Inspired and \textsc{Pearl} as Table.~\ref{tab:conversation-level quality}. 
Given the widespread adoption and proven effectiveness of LLMs in various evaluation tasks \cite{liu2023evaluate,desmond2024evalullm}, we select the LLM-based method for conversational-level evaluation. 
Furthermore, utilizing LLM evaluation offers a distinct advantage: given the unique content and structural characteristics of \textsc{Muse}, conventional manual assessment for conversation-level quality could potentially introduce subjective biases. 
200 conversations are randomly sampled from each dataset, and LLMs are employed to evaluate them on a scale of 0-2 across five dimensions: dialogue naturalness (Natural), logical coherence (Logical), information richness (Informative), product-context relevance (P-C Correlation), and image-text alignment (I-T Correspondence, specifically for multimodal datasets). 
Details are presented in Appendix ~\ref{appendix:conversation-level}. 
The results demonstrate \textsc{Muse}'s superior performance across all five metrics. 
\textsc{Muse}'s high scores in naturalness and coherence establish the fundamental quality. 
The high informative score observed in \textsc{Muse} and \textsc{Pearl} reflects a characteristic advantage of LLM-synthesized datasets. 
The strong P-C Correlation of \textsc{Muse} confirms its suitability for recommendation tasks. 
Regarding I-T Correspondence, \textsc{Muse} outperforms the typical multimodal dialogue dataset MMCONV \cite{liaoC21mmconv}, attributed to its richer integration of image-explanatory elements.  

\subsection{Utterance-level Evaluation}
\label{sec:utterance-level}
In order to evaluate quality at the utterance-level, we randomly select conversation contexts from \textsc{Muse} and instruct annotators to generate responses based on these contexts. 
This process created "artificial utterances" for direct comparison with the original ones within the same contexts.  
Noting that by holding the same contexts and focusing solely on the quality of single-utterance responses, potential biases are significantly reduced. 
Therefore, we choose to apply manual judgment to perform an utterance-level evaluation to assess the quality of conversations in \textsc{Muse} from more perspectives. 
Specifically, annotators are presented with masked paired responses, informed of the given context, and asked to perform anonymous 1:1 comparisons to determine which response is better across four perspectives: Logical, Informative, Natural, and Coherence (Context Coherence). 
Figure.~\ref{fig:utterance-level} demonstrates that original dialogues in \textsc{Muse} are superior quality compared to human-authored dialogues. 
Our observations suggest that this disparity arises from the LLM's ability to thoroughly interpret and present product attributes. 
All annotators are graduate students from our university with expertise in conversational recommendation tasks. 
We provide detailed task descriptions and a fair, anonymous evaluation environment as Appendix \ref{appendix:utterance-level}. 
Three of them are responsible for utterance generation, while the other three conduct anonymous 1:1 evaluations. 

\subsection{Evaluation on Conversational Recommendation Task: Recommend}
\label{sec:eval_rec}
We investigate the applicability of \textsc{Muse} for recommendation tasks by conducting recommendation experiments with \textsc{Muse} on three open-source MLLMs: \texttt{Qwen2-VL-7B-Instruct}\cite{Qwen2VL}, \texttt{LlaVA-Next-LLaMA-8B}\cite{li2024llavanext-strong}, and \texttt{Yi-VL-6B}\cite{young2024yi}. 
Models are tasked with generating queries based on multimodal contexts to recall items for recommendation in the current round. 
Then we use recall@n and mrr@n to evaluate the accuracy. 
The fine-tuned models underwent Low-rank adaptation Finetune (LoRA) \cite{hu2021lora} using 200 conversations as the setting in \cite{liang2024llmredial}, with actual queries serving as the golden responses for training.  
Table.~\ref{tab:model_recommendation} contrasts the performance metrics. 
The consistent performance gains across all models validate our dataset's effectiveness, supporting previous findings \cite{bao2023tallrec,bao2023bi} that suggest LLMs require fine-tuning for recommendation tasks. 
Notably, the relative performance ranking among the three models remained consistent across both two settings, aligning with their respective rankings on OpenCompass's MMBench \cite{liu2025mmbench}. 
This consistency validates our dataset's internal coherence of recommendation logic and demonstrates its discriminative power in differentiating model capabilities. 
\begin{table}
\centering
\caption{Recommendation performance of different models under zero-shot and fine-tuning settings}
\vspace{-2mm}
\resizebox{0.4\textwidth}{!}{
\begin{tabular}{lcccc}
\toprule
\textbf{Setting} & \textbf{Recall@10} & \textbf{Recall@20} & \textbf{MRR@10} & \textbf{MRR@20} \\
\midrule
\multicolumn{5}{l}{\textit{\textbf{\textbf{LlaVA-NEXT-Llama3-8B} }}} \\
\midrule
 Zero-Shot & 0.16 & 0.25 & 0.07 & 0.09 \\
 Finetune & 0.25 & 0.37 & 0.13 & 0.16 \\
\midrule
\multicolumn{5}{l}{\textit{\textbf{Yi-VL-6B}}} \\
\midrule
Zero-Shot & 0.15 & 0.23 & 0.07 & 0.08 \\
Finetune & 0.25 & 0.35 & 0.12 & 0.14 \\
\midrule
\multicolumn{5}{l}{\textit{\textbf{Qwen-2-VL-7B}}} \\
\midrule
Zero-Shot & 0.20 & 0.33 & 0.12 & 0.13 \\
Finetune & \textbf{0.34} & \textbf{0.45} & \textbf{0.22} & \textbf{0.24} \\
\bottomrule
\end{tabular}
}
\vspace{-6mm}
\label{tab:model_recommendation}
\end{table}

\subsection{Evaluation on Conversational Recommendation Task: Response}
Although LLMs excel at general conversation, their response efficacy as specialized rec-assistants warrants investigation. 
We evaluate three MLLMs in Section \ref{sec:eval_rec}, comparing their performance before and after fine-tuning. 
The fine-tuning protocol implements a dual-mode approach: 
(1) For recommendation rounds, the model receives both contextual information and specific multimodal details of the product to be recommended in the current round to generate an appropriate recommendation response. 
(2) For standard conversation rounds, the model generates responses based solely on the dialogue context. 
As illustrated in Table.~\ref{tab:model_response}, fine-tuning significantly improved the alignment between LLM responses and our dataset patterns. 
This improvement demonstrates two key findings: first, our dataset contains learnable response patterns for rec-assistant; second, the enhanced response diversity indicates that LLMs can generate more specific and varied outputs for conversational recommendation after training. 

Additionally, we randomly sample 100 pairs of zero-shot and finetuned responses from different models and combine them for anonymous manual evaluation as Appendix \ref{appendix:utterance-level}.
As shown in the Table.\ref{tab:win_ratio}, responses generated from fine-tuned models are more favored by users. 
Interview feedback indicated that evaluators generally find fine-tuned responses better capture users' key interests. 

\begin{table}
\centering
\caption{Response performance of different models under zero-shot and fine-tuning settings (p-value).}
\vspace{-2mm}
\resizebox{0.48\textwidth}{!}{
\begin{tabular}{lccccc}
\toprule
\textbf{Setting} & \textbf{BLEU-4} & \textbf{ROUGE-1} & \textbf{ROUGE-L} & \textbf{Distinct-4} & \textbf{Avg. Words} \\
\midrule
\multicolumn{6}{l}{\textit{\textbf{LlaVA-NEXT-Llama3-8B}}} \\
\midrule
Zero-Shot & 17.1 & 17.2 & 9.61 & 0.63 & 86.6\\
Finetune & 44.0 & 37.8 & 27.2 & 0.64 & 52.6 \\
\midrule
\multicolumn{6}{l}{\textit{\textbf{Yi-VL-6B}}} \\
\midrule
Zero-Shot & 16.5 & 16.7 & 9.48 & 0.58 & 107.7 \\
Finetune & 44.1 & 37.7 & 27.1 & 0.63 & 54.9 \\
\midrule
\multicolumn{6}{l}{\textit{\textbf{Qwen-2-VL-7B}}} \\
\midrule
Zero-Shot & 41.1 & 35.3 & 23.5 & 0.69 & 72.3 \\
Finetune & \textbf{46.8} & \textbf{42.7} & \textbf{31.5} & \textbf{0.71} & \textbf{48.7} \\
\bottomrule
\end{tabular}
}
\vspace{-2mm}
\label{tab:model_response}
\end{table}

\begin{table}
\centering
\caption{Comparison between responses from zero-shot and finetuned models}
\vspace{-2mm}
\label{tab:win_ratio}
\setlength{\tabcolsep}{20pt}  % 调整列间距，可以根据需要调整数值
\resizebox{0.3\textwidth}{!}{
\begin{tabular}{lcc}
\toprule
& \textbf{Zero-Shot} & \textbf{Finetune} \\
\midrule
\textbf{Win Ratio} & 0.12 & 0.88 \\
\bottomrule
\end{tabular}
}
\vspace{-6mm}
\end{table}

\section{Conclusion}
Existing conversational recommendation (CR) research focuses solely on text, leaving a gap with real-world applications. 
\textsc{Muse}, the first multimodal conversational recommendation dataset with 7,000 Clothing-related conversations, is introduced to bridge the gap.
Validated by LLMs and humans, the conversations in \textsc{Muse} are shown to be highly informative, fluent, and logically coherent. 
Through benchmark testing on several multimodal LLMs (MLLMs), we demonstrate that \textsc{Muse} exhibits reliable recommendation and response logic, making it a valuable resource for CR research. 
The conversations in \textsc{Muse} are automatically generated using a multi-agent framework powered by MLLMs, which leverages a scenario-grounded approach to create user profiles tailored to specific products and simulate realistic CR conversations. 
Addressing the scalability limitations of existing CR data synthesis methods, it holds the potential to expand MCR datasets to include a wider range of domains, users, and products in the future. 

\section{Limitation}
The synthesis process of conversations in \textsc{Muse} relies extensively on the powerful capabilities of multimodal large models (MLLMs). 
As a result, the data quality is inherently influenced by the model's capabilities.
Due to cost constraints, we opt to use \texttt{gpt-4o-mini} as the primary model instead of the more powerful but more expensive \texttt{gpt-4o}. 
Similarly, because API calls for image processing are expensive and each conversation synthesis involves reading a large number of images, we are unable to scale the dataset to the size of pure text datasets like \textsc{Pearl} and LLM-Redial. 
In addition, while \textsc{Muse}'s conversations already include more rounds than some existing datasets and can be further extended based on specific settings, increasing the context length and the number of images can impact the response generation capabilities of LLMs. 
As a result, we do not pursue the synthesis of CR data with ultra-long dialogue rounds. In the future, we plan to explore prompt compression techniques to address this limitation. 
% \bibliography{ref.bib}

\appendix
\section{Implementation Details}
\subsection{Data Preprocess}
\label{appendix:data_preprocess}
During data preprocessing, our primary task is to build a local database using the multimodal information of the products, which supports subsequent product retrieval. 
The local product database contains a subset of the Amazon Cloth, Shoes, and Jewelry products. 
The process is as follows. 
Due to the extensive volume of products in the initial dataset, we strategically reduce the product count and eliminate items with incomplete multimodal information to yield a refined dataset comprising 94,209 products. 
In the subsequent phase, we construct a local product database utilizing both the visual and text information of the product. 
Initially, we utilize MLLMs (\texttt{gpt-4o-mini}) to summarize the product's multimodal data, eliminating redundant details, marketing language, and other noise. 
This process ensures that the revised summary focuses on the fundamental and visual attributes of the product itself. 
Following this, we use the summary of products to establish a local product database with BGE-M3\cite{multim3}. 
Alternatively, multimodal embedding models can be employed, which hold the advantage of retaining more of the original information; however, they also present the risk of introducing additional noise. 
Finally, 13,754 products in the local product database are used for the synthesis of \textsc{Muse} data.

\begin{figure}
    \centering
    \includegraphics[width=0.98\linewidth]{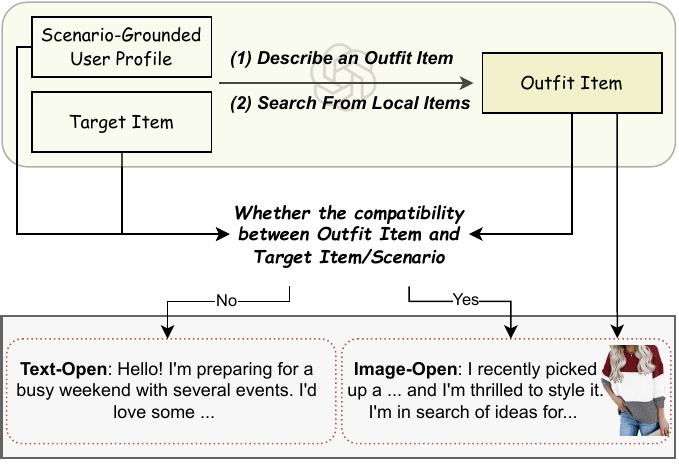}
    \caption{Generation of multimodal-open Open-Dialogues.}
    \label{fig:image_open}
\end{figure}

\begin{figure*}
    \centering
    \includegraphics[width=0.98\linewidth]{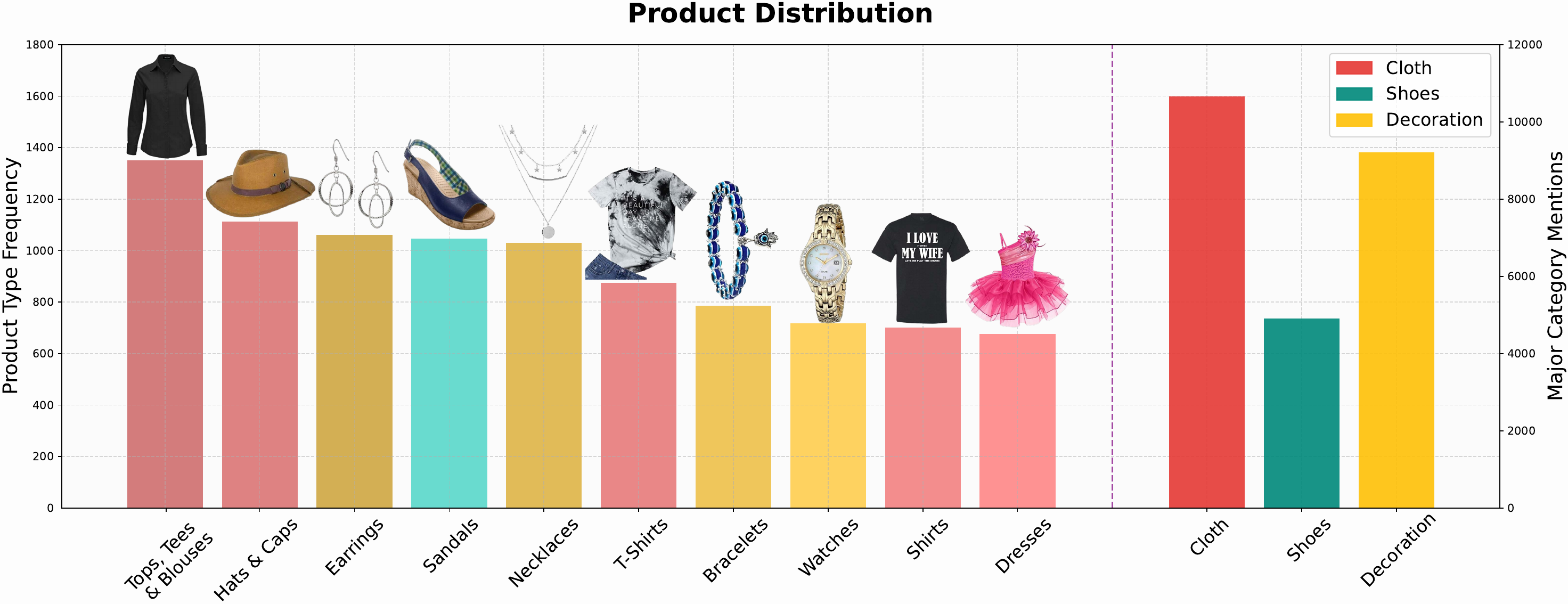}
    \caption{Distribution of mentioned products in \textsc{Muse}. }
    \label{fig:product_distribution}
\end{figure*}

\subsection{Product Distribution}
\label{appendix:product}
To gain a deeper understanding of our dataset's composition, we visualize all the mentioned product types, as shown in the Figure.~\ref{fig:product_distribution}.
On the right side of the figure, we display the proportions of the three main categories classified at the broadest level: Clothing, Shoes, and Decoration. 
(Note: The original dataset labels the third category as Jewelry, but we rename it Decoration after observing that items like wallets are also grouped into this category.) 
Meanwhile, the left side of the figure illustrates the frequency of the top-10 product types. 
\subsection{Details of the Open-dialogues}
\label{appendix:open-dialogue}
Our dataset incorporates two distinct approaches for conversation initiation. 
The first approach follows traditional crowdsourcing conversational recommendation systems, where users articulate their needs through text, which is 'text-open.` 
To integrate multimodal elements, we introduce a second approach that allows users to express their requirements through images, which is 'multimodal-open`. 
Given the characteristics of our local product clothing dataset, we specifically align multimodal-open with the outfit problem. 
We frame an multimodal-open case as follows: a user who has already purchased an item of clothing (outfit item) seeks recommendations for a complementary item (target item) that both coordinates with the outfit item and aligns with the scenario. 
In this case, the user not only articulates the requirements verbally but also shares an image of the outfit item with the rec-assistant to ensure visual compatibility. 

The workflow of the multimodal-open case can be seen in Figure.~\ref{fig:image_open}.
Specifically, an MLLM (\texttt{gpt-4o-mini}) is employed to analyze outfit requirements based on the user's target product/scenario, automatically generating descriptions of potential matching items. 
These descriptions are served as queries to search our product database. 
Since the retrieved local product may not perfectly match these generated descriptions - a commom occurence - the model performs an additional compatibility assessment. 
The secondary evaluation determines whether the retrieved local products can effectively coordinate with the user's target item while fulfilling the intended scenario requirements. 
When the secondary evaluation is successful (with an approximate success rate of 47\%), the conversation adopts an multimodal-open format, in which users express both their scenario-related requirements and outfit-item-driven needs, allowing the rec-assistant to identify suitable recommendations. 

\subsection{Architecture of Rewriter}
\label{appendix:rewriter_structure}
\begin{figure}
    \centering
    \includegraphics[width=0.48\textwidth]{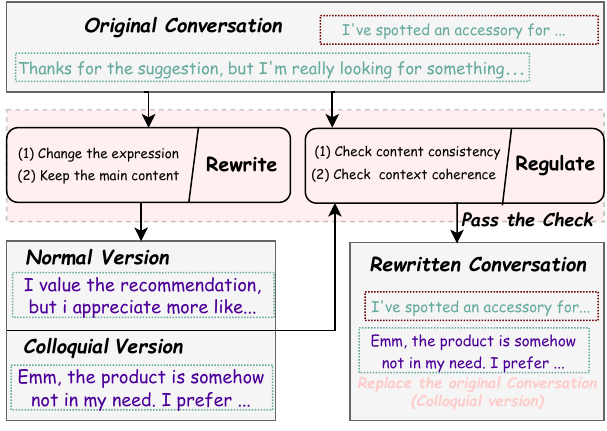}
    \caption{The basic workflow of Rewriter.}
    \label{fig:rewriter}
\end{figure}

During the simulated conversation generation, we employ a low-temperature setting for the MLLM to maintain process stability, though this constrains the diversity of expressions and sentence structures. 
To address this limitation, we develop a Rewriter system, as illustrated in Figure.~\ref{fig:rewriter}. 
The system operates in two phases: first, using the LLM at high temperature to restructure and rephrase the original conversation, then supervising the output at low temperature to preserve semantic consistency with the source dialogue. 
Additionally, we probabilistically inject prompts that encourage the use of colloquial expressions, resulting in more authentic, conversation-like exchanges. 
This approach effectively brings diversity to sentences and keeps semantic fidelity. 

\subsection{Multimodal Large Language Model Settings for Synthesizing Data}
\label{appendix:mmsetting}
Our data synthesis methodology fundamentally leverages the advanced capabilities of LLMs, necessitating the selection of a robust and reliable model as our foundation. 
Given our computational resource constraints, we adopt an API-based deployment strategy. 
After evaluating the performance-cost trade-offs, we select \texttt{gpt-4o-mini} as our primary Multi-modal Large Language Model (MLLM) foundation. 
Within our multi-agent framework for Scenario-Grounded User Profile Generator and Simulated Conversation Generator, we set the temperature parameter of \texttt{gpt-4o-mini} to 0.1-0.2 for agent role-playing. 
This low-temperature configuration, coupled with the model's robust instruction-following capabilities, ensures consistent and reliable agent behavior. 
For the Rewriter, we implement \texttt{Claude-3.5-haiku}\footnote{https://claude.ai} as the foundational model. 
This decision stems from our empirical observations that \texttt{gpt-4o-mini} exhibited limited syntactic diversity and lexical richness in dialogue generation. 
Even adjusting the temperature to higher values (0.8-0.9) failed to enhance output variability. 
And the validation operation in the Rewriter is based on \texttt{gpt-4o-mini}. 

\section{Further Analysis}
\subsection{Discussion on Data Quality}
\begin{figure}
    \centering
    \includegraphics[width=0.95\linewidth]{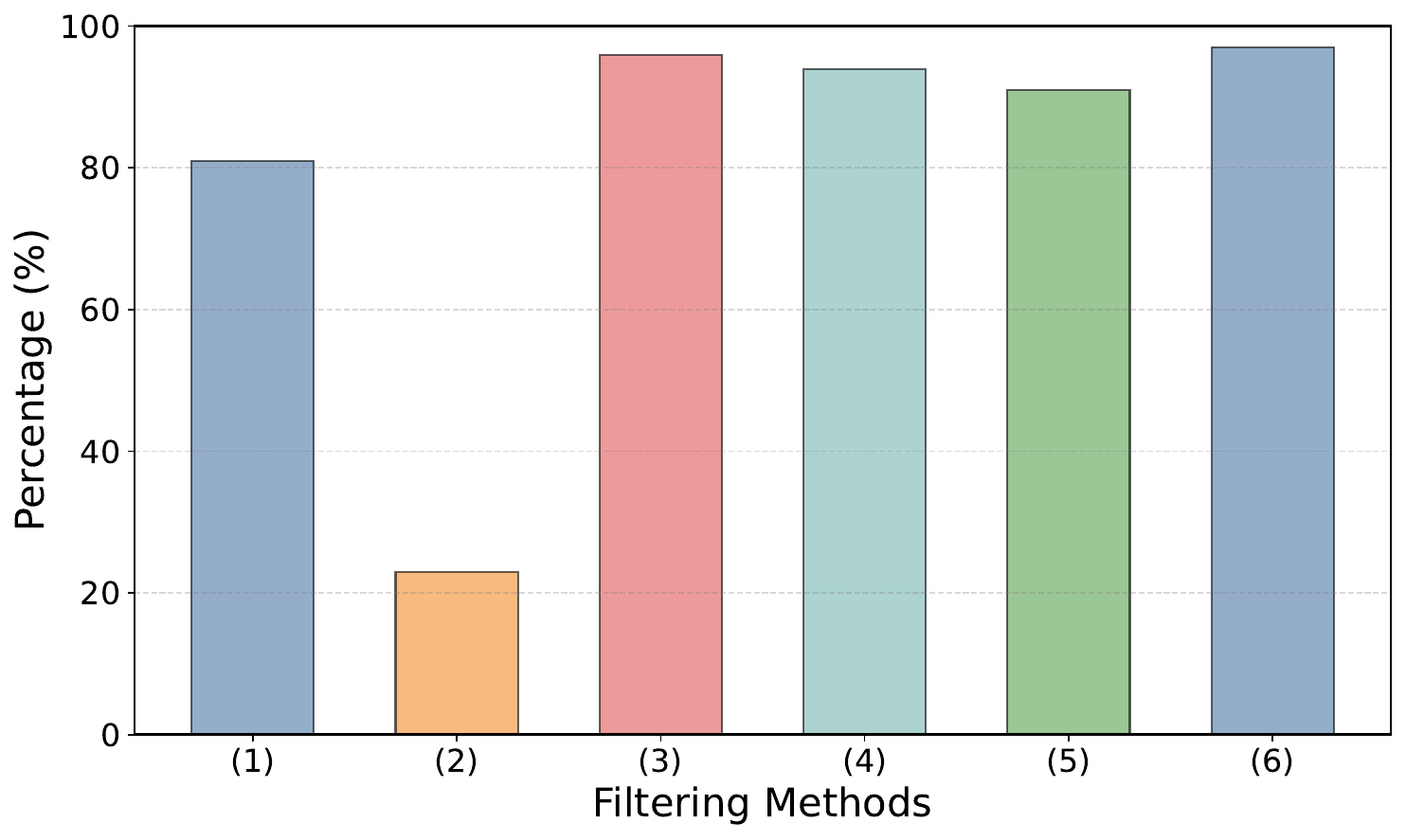}
    \caption{Comparison of pass probability of the six filtering methods.}
    \label{fig:quality_percentage}
\end{figure}

Our framework actually has implemented five automatic screenings to ensure the stability of quality and diversity: 
(1) BLEU deduplication for basic scenarios. 
(2) Quality screening for the matching rationality of users, scenarios, and products. 
(3) BLEU deduplication for the generated purchase backstory. 
(4) Content consistency screening for the rewritten dialogue in the Rewriter. 
(5) Quality screening for a whole conversation by LLM. 
Considering the evaluation of the overall reliability of the public dataset, we also use (6) manual screening, but it turn out that after the previous five screenings, the quality of the retained conversations was already quite guaranteed. 
Manual screening only filters out a small number of unqualified conversations discussed in the experimental section. 
Figure.~\ref{fig:quality_percentage} indicates the probability of passing different screening measures. 

\subsection{Extended Manual Verification}
In the main text, we use manual verification to perform utterance-level comparisons, where humans generate responses based on the same context and compare them with the responses in \textsc{Muse}. 
To provide more comprehensive manual comparisons, we introduce an additional comparison experiment here. 
In this experiment, we collect more human-generated replies and replace selected responses in the existing \textsc{Muse} conversations. 
Specifically, we randomly select 50 complete conversations from \textsc{Muse} and, for each conversation, replace 4 responses (two from users and two from the rec-assistant) with human-generated replies to create a conversation-level variant. 
Another user then judges which conversation—original or modified—is better and smoother overall. 
The results, shown in the Table.~\ref{tab:muse_comparison_conversation}, show that even under these replacement conditions, the original conversations in \textsc{Muse} are still preferred by human evaluators.
\begin{table}
    \centering
    \resizebox{0.4\textwidth}{!}{
    \setlength{\tabcolsep}{30pt}  % 调整列间距
    \begin{tabular}{lcc}
        \toprule
        \textbf{Data} & \textbf{Muse} & \textbf{Muse (Replace)} \\
        \midrule
        Win Rate & 0.68 & 0.32 \\
        \bottomrule
    \end{tabular}
    }
    \caption{Comparison of Win Rates between Original Muse and Response-replaced Muse}
    \label{tab:muse_comparison_conversation}
\end{table}

\subsection{Cost Analysis}
The multi-agent framework behind \textsc{Muse} consists of four steps. The first step focuses on basic scene expansion, utilizing \texttt{gpt-4o-mini} to generate approximately 500 diverse real-world scenarios. 
Given the concise nature of inputs and outputs in this step, the associated costs are negligible. 
While employing more advanced models like \texttt{gpt-4o} or \texttt{claude-3.5-haiku/sonnet} would increase expenses, they could potentially yield more diverse and nuanced scenes.
The second step involves user-scenario-product matching to generate scenario-grounded user profiles, also powered by \texttt{gpt-4o-mini}. 
The input consists of a single image and some text and the output is not long.  
However, due to stringent quality control measures—including user-scenario-product alignment verification and BLEU-based deduplication—only 21.6\% of generated user profiles meet the qualification standards. 
As a result, the average cost for producing a qualified user profile is approximately \$0.009, though this can be reduced to \$0.001 by utilizing product image descriptions as alternative inputs. 
The third component implements iterative conversation generation using \texttt{gpt-4o-mini}, which requires multiple image/text readings.
The generation cost per complete \textsc{Muse} conversation amounts to roughly \$0.033. 
The final step encompasses conversation rewrite and quality assessment, as detailed in Appendix \ref{appendix:mmsetting}. 
For conversation rewriting, we employ \texttt{claude-3.5-haiku} with high temperature settings. 
Since this phase processes plain text for both input and output, and considering the pass rate along with \texttt{gpt-4o-mini}'s quality supervision costs, the average expense for refining a complete conversation is \$0.011. 
The dialogue quality review phase evaluates content and outputs scores at \$0.008 per review. 
However, since image-text alignment is already verified during generation, we can optimize costs by screening text only, reducing the expense to \$0.001 per review while maintaining quality standards. 

\subsection{Discussions of N-gram Diversity Patterns}
\label{appendix:n_gram}
The aforementioned analysis reveals that the Distinct-4 scores are significantly low in both the Pearl and Muse (without the Rewriter).  
This phenomenon can be attributed to the behavior of GPT-series models operating at low temperatures. 
While these models demonstrate strong instruction-following capabilities, they tend to generate dialogues with highly repetitive sentence patterns. 
To provide concrete evidence of this pattern repetition, we conduct a detailed analysis of the most frequent 4-grams in each dataset, which offers a clear visualization of this linguistic homogeneity. 

\begin{figure*}
    \centering
    \includegraphics[width=0.96\textwidth]{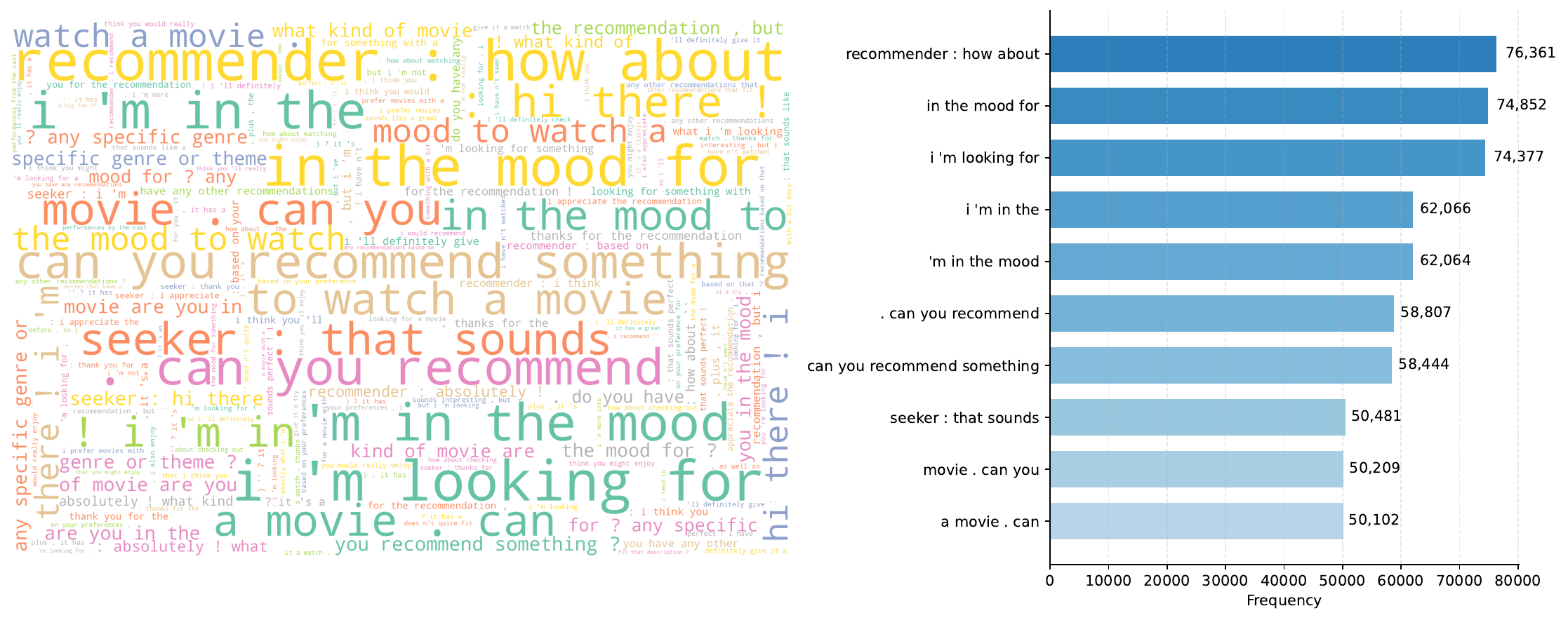}
    \caption{Word-cloud of top-200 4-grams in Pearl (left) and top-10 4-gram's frequency in Pearl (Right)}
    \label{fig:top_200_original_muse}
\end{figure*}

\begin{figure*}
    \centering
    \includegraphics[width=0.96\textwidth]{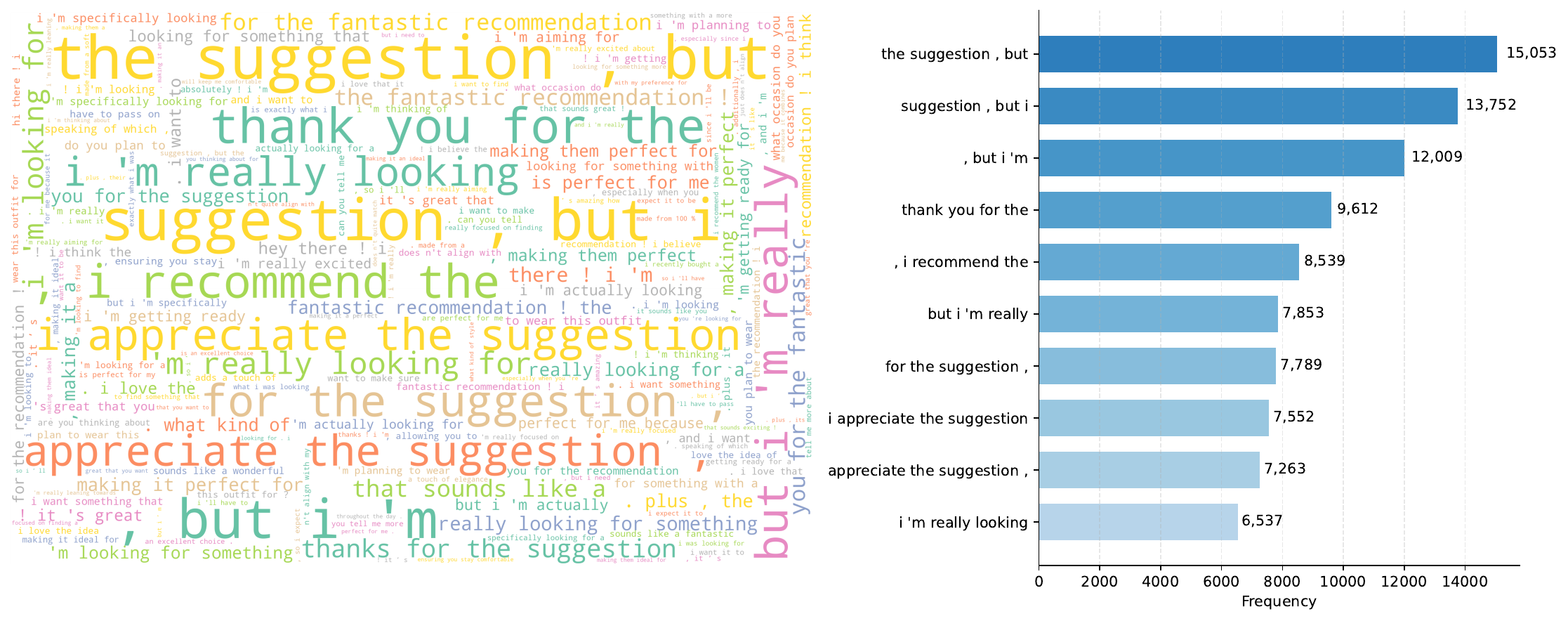}
    \caption{Word-cloud of top-200 4-grams in Muse without Rewriter (left) and top-10 4-gram's frequency in Muse without Rewriter (Right)}
    \label{fig:top_200_pearl}
\end{figure*}

\begin{figure*}
    \centering
    \includegraphics[width=0.96\textwidth]{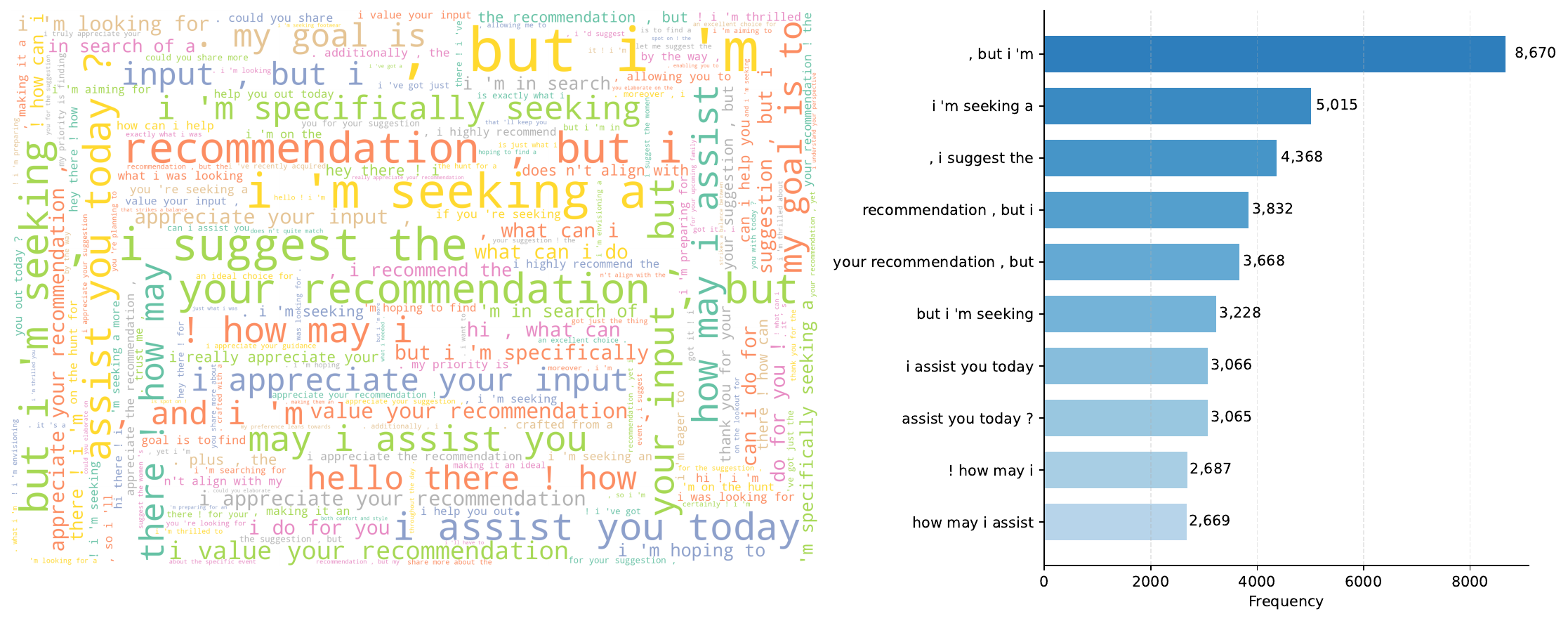}
    \caption{Word-cloud of top-200 4-grams in Muse with Rewriter (left) and top-10 4-gram's frequency in Muse with Rewriter (Right)}
    \label{fig:top_200_rewrite_muse}
\end{figure*}

Figure.~\ref{fig:top_200_original_muse}, \ref{fig:top_200_pearl}, and \ref{fig:top_200_rewrite_muse} present a comprehensive visualization of the 4-gram distribution in each dataset. 
The left panels display word clouds of the top 200 4-gram phrases, while the right panels show frequency distributions of the top 10 most recurring phrases. 
The word cloud visualizations clearly demonstrate that the majority of these 4-grams consist of repetitive sentence structures and fixed phrasal patterns. 
It substantiates our hypothesis that the low Distinct-4 scores are a direct result of this limited linguistic variability, as the prevalence of standardized sentence constructions inherently reduces 4-gram diversity. 
The analysis reveals a notable disparity in phrase frequency distributions. 
In both Pearl and Muse (without the Rewriter), the top 10 4-gram phrases exhibit disproportionately high frequencies relative to the total utterance count. 
However, following the application of Rewriter, we observe a substantial reduction in the frequency of these recurring phrases in Muse. 
This decline in repetitive patterns, coupled with the increased 4-gram diversity shown in Table 1, provides compelling evidence that the Rewriter successfully enhances linguistic variability by generating more diverse and sophisticated sentence structures. 
Additional diversity metrics are shown in Table.~\ref{tab:more_diversity_metrics}. 

\begin{table}[htbp]
\centering
\resizebox{0.48\textwidth}{!}{
\begin{tabular}{lrrrrr}
\hline
\textbf{Metric} & \textbf{Redial} & \textbf{MMCONV} & \textbf{PEARL} & \textbf{Muse*} & \textbf{Muse} \\ 
\hline
Distinct-2 & 0.09 & 0.08 & 0.03 & 0.06 & 0.09 \\
Distinct-3 & 0.27 & 0.23 & 0.09 & 0.19 & 0.30 \\
Distinct-4 & 0.46 & 0.39 & 0.18 & 0.37 & 0.54 \\
\midrule
2-gram Specificity & 8.63 & 6.10 & 1.81 & 5.62 & 8.09 \\
3-gram Specificity & 24.30 & 17.50 & 6.77 & 18.44 & 27.20 \\
4-gram Specificity & 41.10 & 29.90 & 14.75 & 35.06 & 48.90 \\
\hline
\end{tabular}
}
\caption{Lexical diversity metrics across different conversational datasets. Muse* stands for Muse (without Rewriter). }
\label{tab:more_diversity_metrics}
\end{table}

\subsection{Discussion of the Necessity of Images}
While Muse presents both visual and textual product information in all product-related conversations, we investigate whether state-of-the-art multimodal models' caption capabilities could effectively convert visual information into textual descriptions for direct integration into conversations. 
We select 200 conversations, preserving the original textual components and substituting the visual elements with image descriptions generated by \texttt{gpt-4o}. 
We conduct an A/B test comparing human perception of two datasets: the original \textsc{Muse} data (containing images) and the pure text version (with image descriptions). 
The results in Table.~\ref{tab:muse_comparison} decisively favor the original \textsc{Muse} data with images, except in cases where discussions focus solely on product attributes. 
This preference can be attributed to two factors: first, humans process visual information more efficiently than text descriptions; second, even \texttt{gpt-4o}'s image descriptions occasionally contain style interpretation inaccuracies. 
Additionally, text containing numerous visual features significantly increases dialogue length, making it more challenging for users to identify key information quickly. 
\begin{table}
    \centering
    \resizebox{0.4\textwidth}{!}{
    \setlength{\tabcolsep}{30pt}  % 调整列间距
    \begin{tabular}{lcc}
        \toprule
        \textbf{Data} & \textbf{Muse} & \textbf{Muse (Text)} \\
        \midrule
        Win Rate & 0.97 & 0.03 \\
        \bottomrule
    \end{tabular}
    }
    \caption{Comparison of Win Rates between Original Muse and Text-only Muse}
    \label{tab:muse_comparison}
\end{table}

\subsection{Discussion about the Differences between Muse and Existing Multimodal Datasets}
We further elaborate on the differences between \textsc{Muse} and existing multimodal datasets, mainly about SURE and SIMMC \cite{long2023multimodal,crook2021situated,kottur2021simmc}. 
These datasets constrain conversations to specific VR scenarios and focus on multimodal interactions centered around recommendation tasks. 
They primarily emphasize spatial relationships in conversations, such as "Please introduce the red clothes just above the jeans," while providing limited content related to user preferences. Additionally, the products discussed in these conversations are restricted to a small selection available within the VR environment, and these products often lack detailed descriptions. 
The "conversational recommendation task" described in this paper refers to a conversation driven by user needs and interests. 
The goal is to identify products that meet user requirements from a large pool of options, based on the interest feedback provided by the user. 
Each product is accompanied by unique descriptions, including both images and text. This setup aligns with the task definition of traditional recommendation systems, such as collaborative filtering and sequential recommendation tasks, where the objective is to accurately identify products that match user preferences from massive datasets. 
The task scenario in \textsc{Muse} mirrors real-world situations, such as engaging in a conversation with customer service while shopping online. 
Therefore, fundamental differences exist between datasets like \textsc{Muse} and SURE-type VR datasets. 
\textsc{Muse} aligns more closely with the definition of conversational recommendation tasks as outlined in existing datasets \cite{li2018towards,zhou2020topic}. 
\section{Case Study}
\subsection{Scenario-Grounded User Profile Case}
\begin{figure*}
    \centering
    \includegraphics[width=0.8\textwidth]{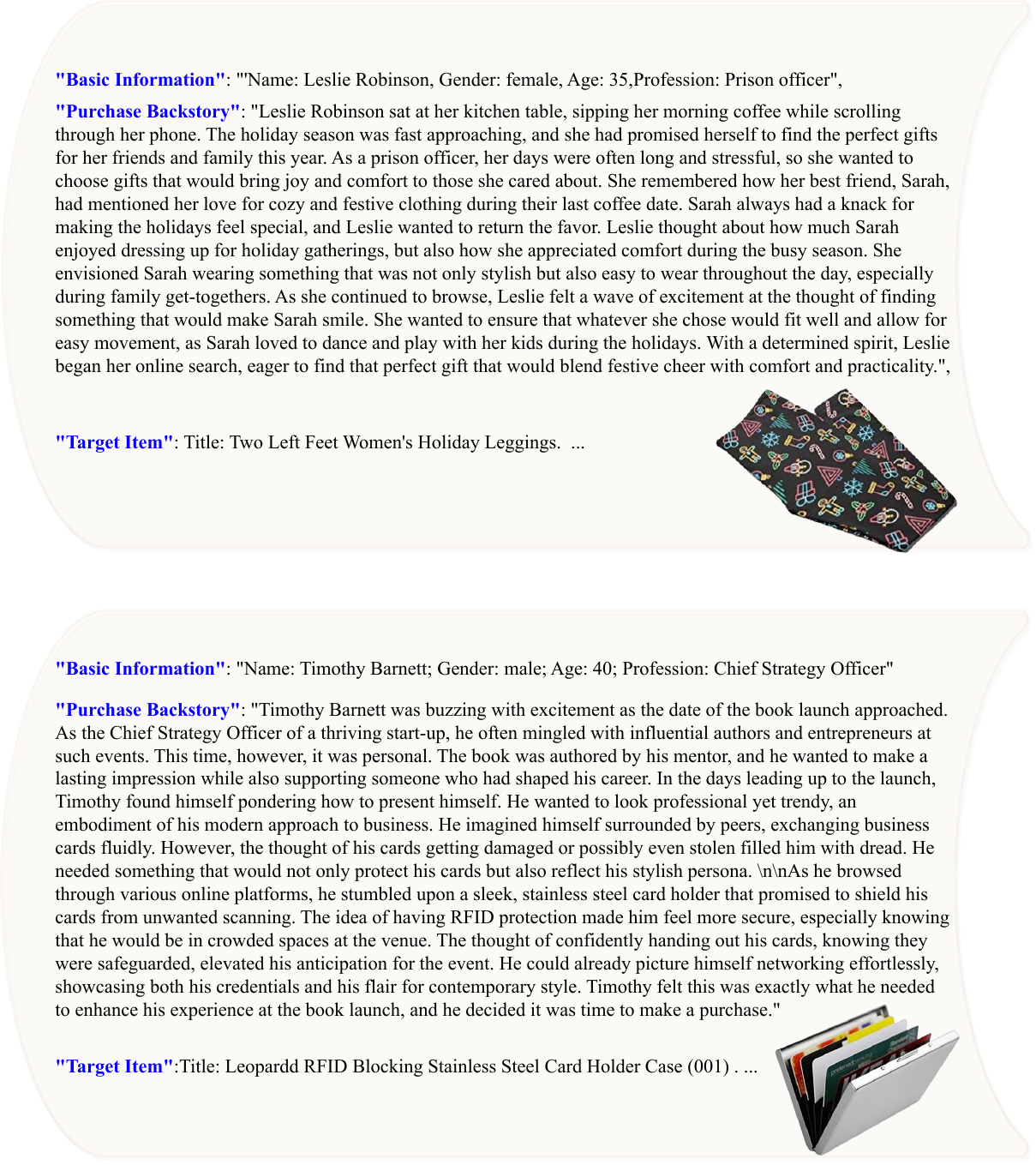}
    \caption{Two cases of the Scenario-Grounded User Profiles}
    \label{fig:case_user_profile}
\end{figure*}

As demonstrated in Figure.~\ref{fig:case_user_profile}, we present two scenario-grounded user profiles that exemplify how individual needs seamlessly align with specific scenarios. 
The inherent connection justifies our decision to incorporate real-world scenarios into the personality generation process. 

\subsection{Conversation Case}
\begin{figure*}
    \centering
    \includegraphics[width=0.8\textwidth]{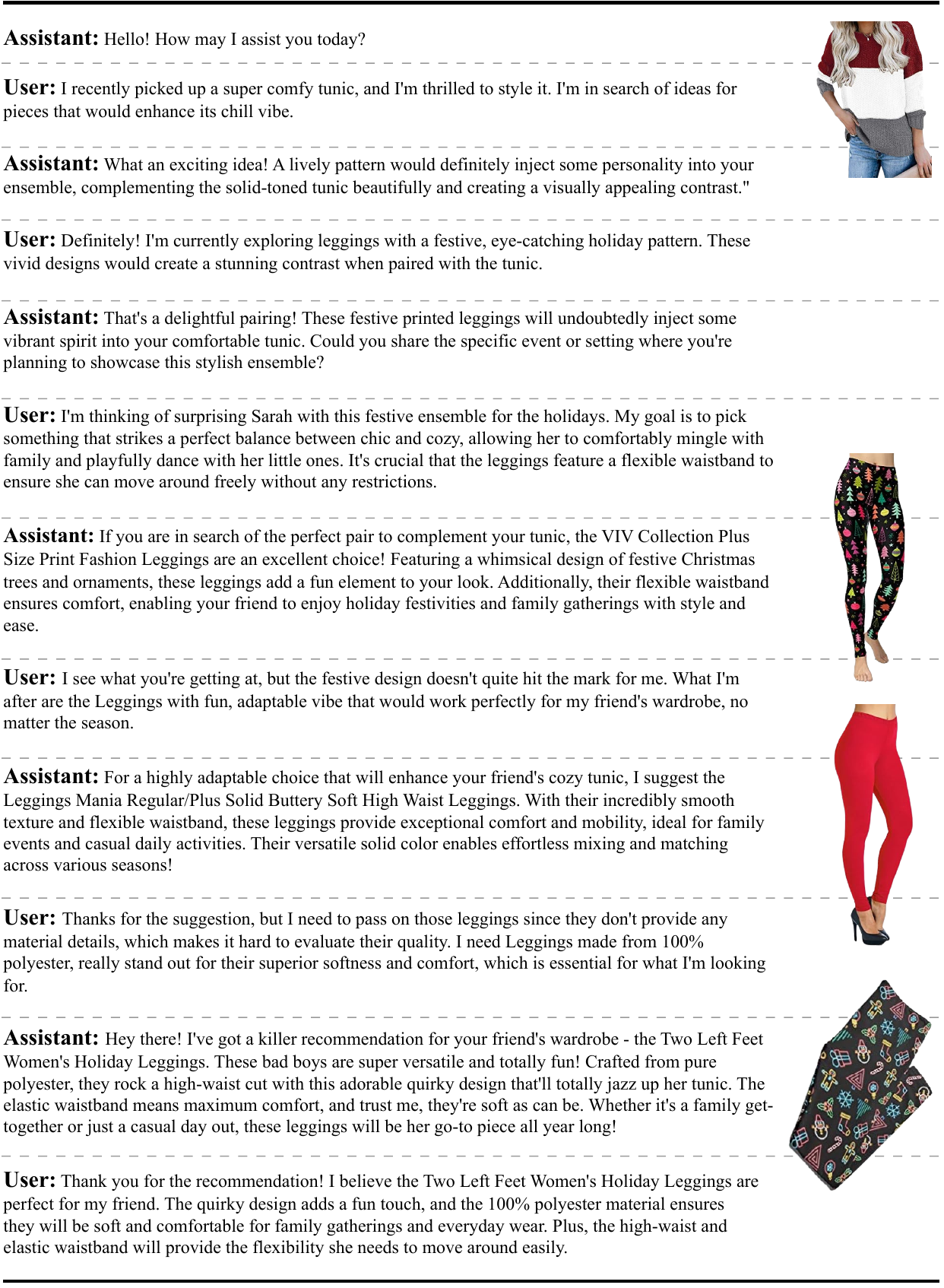}
    \caption{Text-open conversation case.}
    \label{fig:case_conv1}
\end{figure*}

\begin{figure*}
    \centering
    \includegraphics[width=0.7\textwidth]{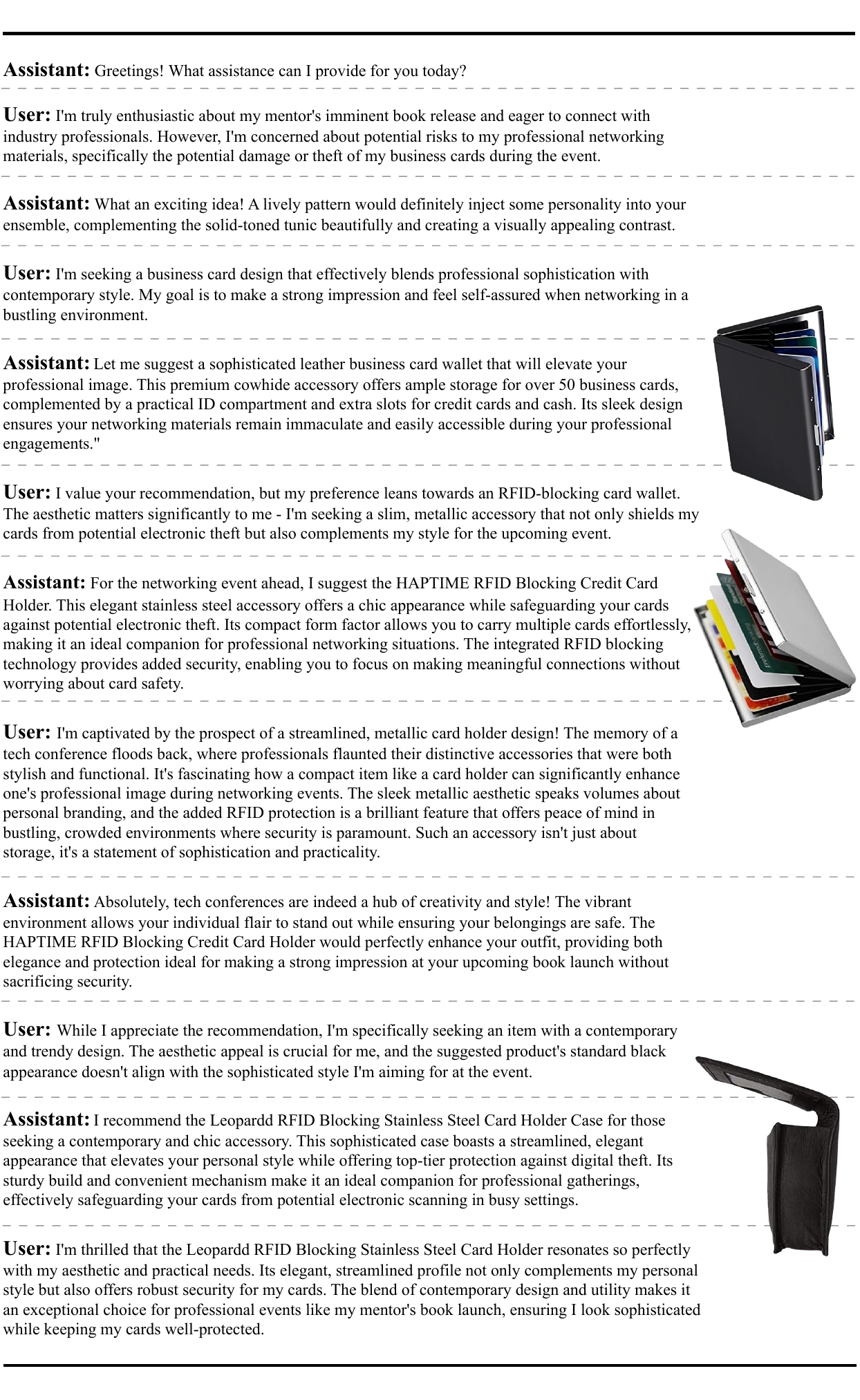}
    \caption{Multimodal-open conversation case.}
    \label{fig:case_conv2}
\end{figure*}

Figure.~\ref{fig:case_conv1} and Figure.~\ref{fig:case_conv2} illustrate two distinct conversational interactions within Muse, each corresponding to the scenarios described above. 
Figure.~\ref{fig:case_conv1} demonstrates a dialogue initiated through the multimodal-open mechanism, while Figure.~\ref{fig:case_conv2} displays a text-open interaction that includes chit-chat. 
The conversation content features comprehensive explainable recommendation factors, encompassing both contextual scene requirements and visual matching criteria. 

\section{Prompt Template and Evaluation Setting}
Compared with Pearl and LLM-Redial, our model incorporates more prompt templates for stricter process control. We present three important prompts here. 
Including scenario-grounded user profile generation, user-side and recommendation assistant-side prompts and the conversation-level evaluation.
For more information, please check \url{https://anonymous.4open.science/r/Muse-0086}. 

\subsection{Scenario-Grounded User Profile Generator}
\begin{figure*}
    \centering
    \includegraphics[width=0.7\textwidth]{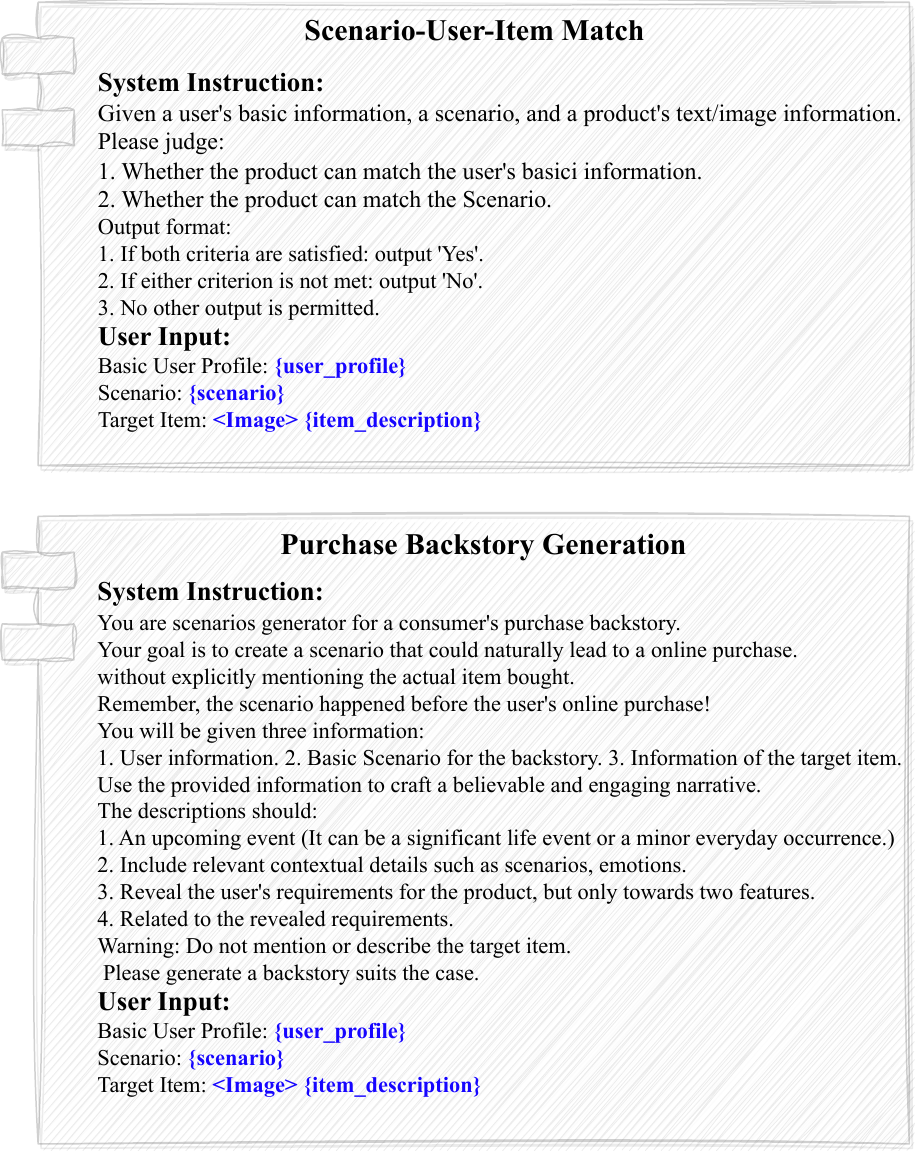}
    \caption{Some prompts of Scenario-Grounded driven User Profile Generator.}
    \label{fig:prompt_scenario}
\end{figure*}

The steps of the Scenario-Grounded User Profile Generator are divided into two steps as in Figure.~ \ref{fig:prompt_scenario}. 
The first step is to screen the matched basic scenario, user basic profile, and target product. 
The second step is to generate the backstory of the user to demonstrate his/her preferences. 

\subsection{Actions of User}
\begin{figure*}
    \centering
    \includegraphics[width=0.7\textwidth]{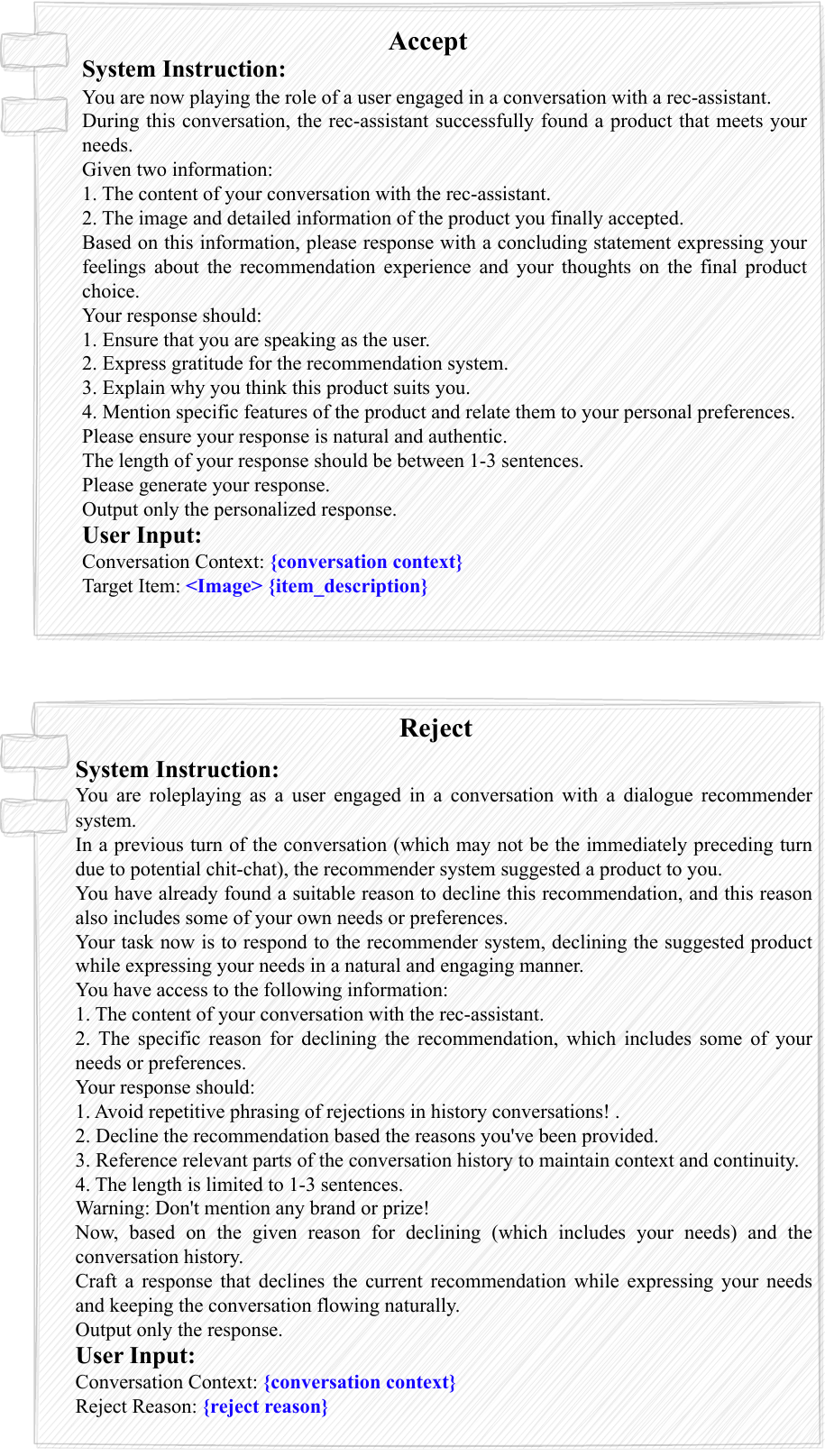}
    \caption{Some prompts of User Simulator.}
    \label{fig:prompt_user}
\end{figure*}

Users exhibit two primary actions: accepting a recommendation and rejecting a recommendation, with their corresponding prompt templates illustrated in Figure.~ \ref{fig:prompt_user}. 
Additionally, users demonstrate two secondary action patterns: chit-chatting and comparing item (find reason for rejecting the item). 
\subsection{Actions of Rec-assistant}
\begin{figure*}
    \centering
    \includegraphics[width=0.7\textwidth]{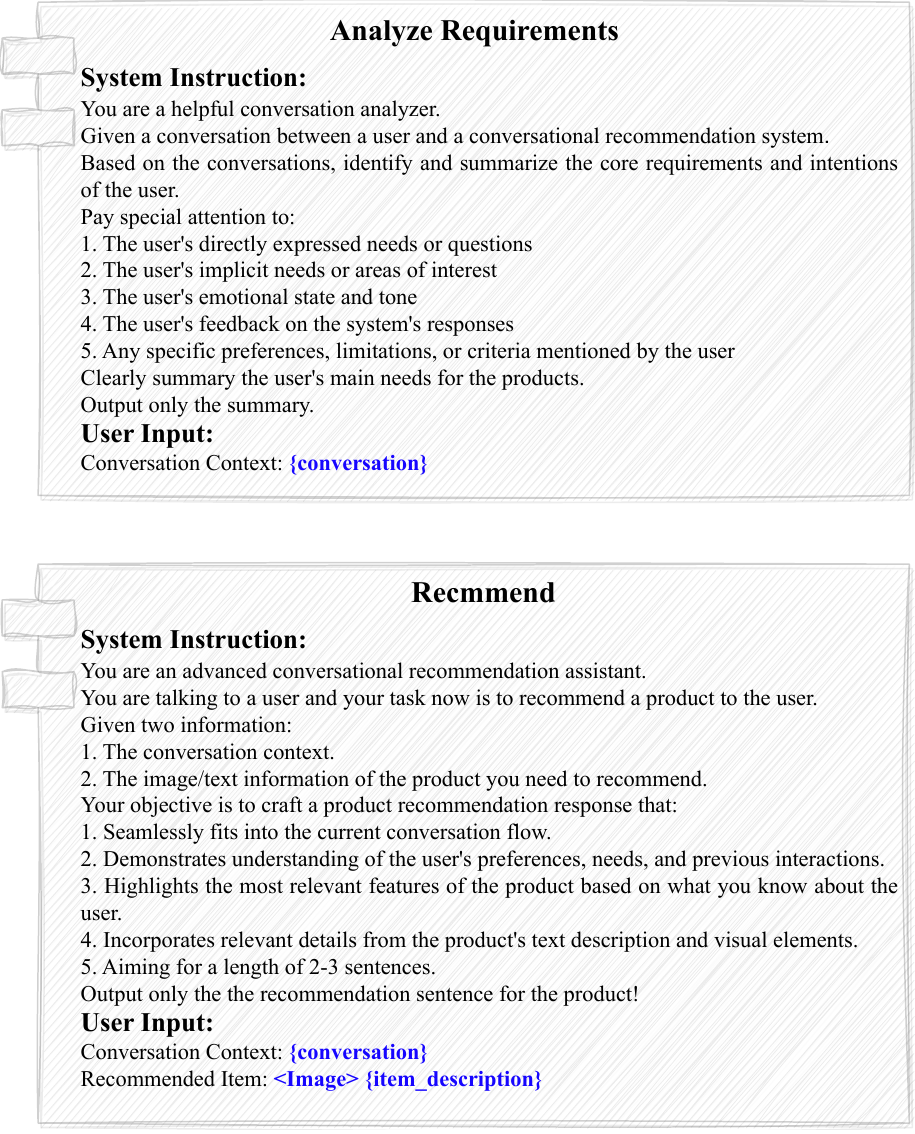}
    \caption{Some prompts of Rec-assistant Simulator.}
    \label{fig:prompt_assistant}
\end{figure*}

The recommendation assistant comprises two components: a Chatter and a Querier. 
Figure.~\ref{fig:prompt_assistant} presents the prompts for two essential functions: the Querier's generation of basic product queries and the Chatter's recommendation process. 

\subsection{Setting for Conversation-level Evaluation}
\label{appendix:conversation-level}
\begin{figure*}
    \centering
    \includegraphics[width=0.7\textwidth]{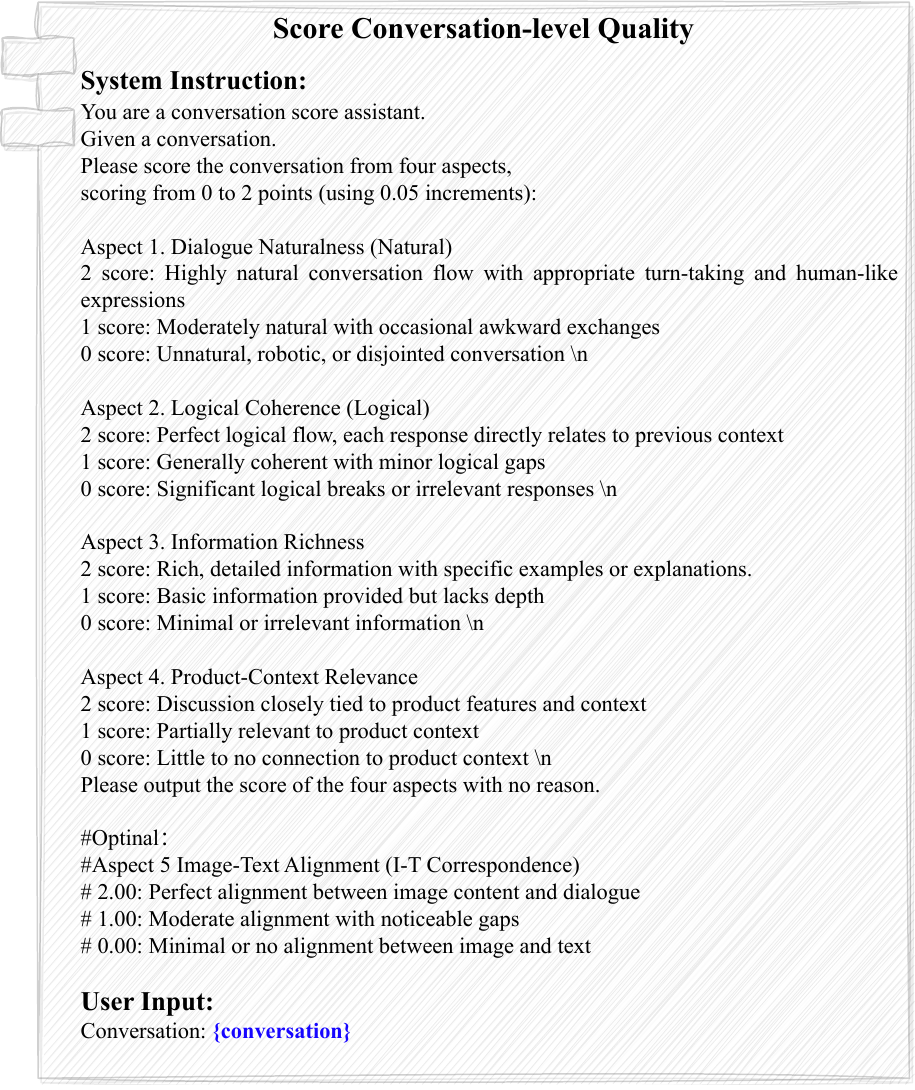}
    \caption{The evaluation prompt for LLM-based assessment of conversation-level quality}
    \label{fig:prompt_evaluation}
\end{figure*}

Here at Figure.~\ref{fig:prompt_evaluation}, we present the evaluation prompt designed to assess conversation-level quality. 
The prompt instructs the large language model to conduct multi-dimensional scoring, providing detailed scoring criteria for each dimension with specific performance benchmarks. 
The quality of a multimodal conversational recommendation dataset can be effectively evaluated using the following five aspects. 
Dialogue Naturalness (Natural) measures how fluent and human-like the conversations are, ensuring realistic and engaging interactions. 
Logical Coherence (Logical) assesses whether responses align logically with previous dialogue turns, maintaining contextual consistency. 
Information Richness (Informative) evaluates the diversity and relevance of details provided, which is critical for generating meaningful and helpful recommendations. 
Product-Context Relevance (P-C Correlation) examines the alignment between recommended products and specific user scenarios, ensuring personalized and context-aware suggestions. Finally, Image-Text Alignment (I-T Correspondence) focuses on the consistency between visual and textual information, crucial for leveraging multimodal data effectively. 
These five aspects comprehensively cover linguistic, contextual, and multimodal dimensions, ensuring the dataset supports realistic, relevant, and high-quality conversational models. 
We employ multiple independent scoring rounds and average the scores to ensure reliability. 

\subsection{Setting for Manual Evaluation}
\begin{figure*}
    \centering
    \includegraphics[width=0.7\textwidth]{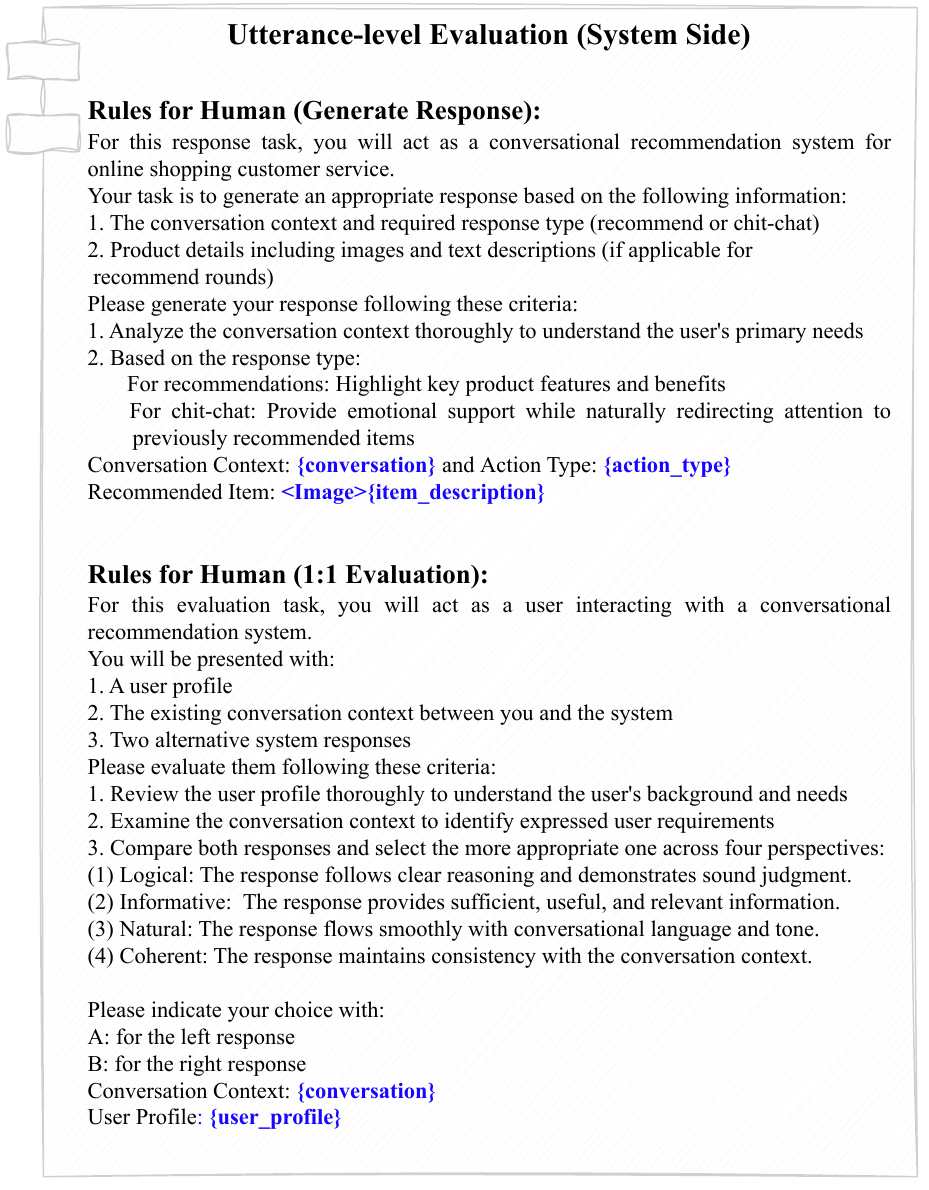}
    \caption{The evaluation rules for manual evaluation of utterance-level quality}
    \label{fig:prompt_human_system}
\end{figure*}

\begin{figure*}
    \centering
    \includegraphics[width=0.7\textwidth]{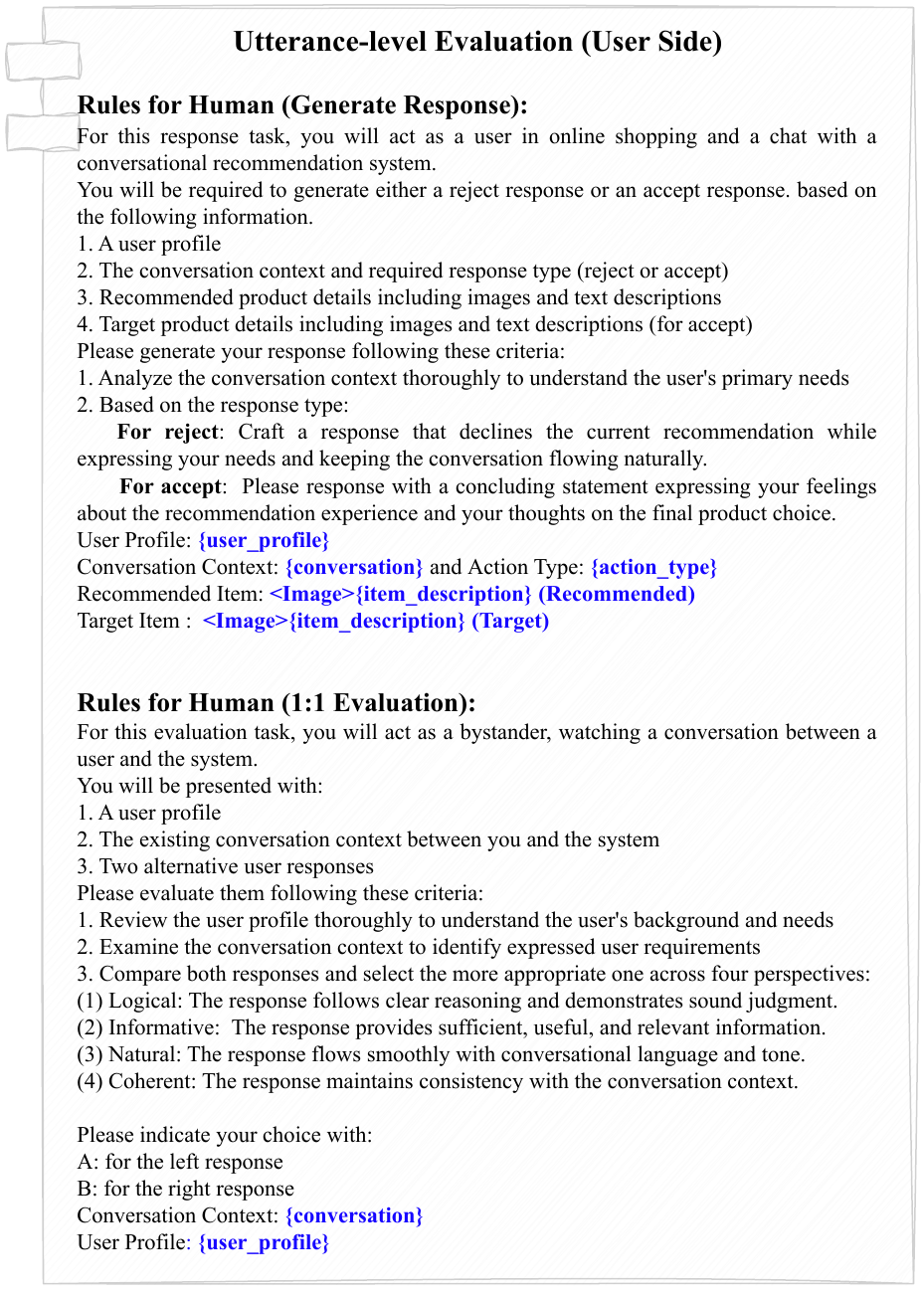}
    \caption{The evaluation rules for manual evaluation of utterance-level quality}
    \label{fig:prompt_human_user}
\end{figure*}

For dialogue quality evaluation at the utterance-level, we employ a two-phase approach as shown in Figure. \ref{fig:prompt_human_system} and Figure. \ref{fig:prompt_human_user}. 
In the first phase, we recruit annotators to generate artificial dialogues based on clear task instructions without any external prompts or interventions. 
In the second phase, we compile a test set containing the dialogue context, newly collected responses, and original dialogues from Muse. 
These are randomly shuffled to ensure unbiased assessment. 
We then conduct blind A/B testing with a separate group of annotators to get final results.

\label{appendix:utterance-level}


\end{document}